\documentclass[10pt]{amsart}
\usepackage{amsmath,amssymb,amsthm,amscd,mathrsfs}
\usepackage{latexsym,amsmath,amssymb,graphicx}
\usepackage[all]{xy}
\usepackage{color}
\usepackage{epsf}
\xyoption{v2}
\xyoption{2cell}
\xyoption{dvips}
\usepackage{youngtab}
\CompileMatrices
\numberwithin{equation}{section}
\newtheorem{prop}{Proposition}[section]
\newtheorem{theo}[prop]{Theorem}
\newtheorem{lemm}[prop]{Lemma}

\newtheorem{exam}[prop]{Example}

\newtheorem{conj}[prop]{Conjecture}

\numberwithin{equation}{section}
\newcommand{\be}{\begin{equation}}
\newcommand{\ee}{\end{equation}}

\newcommand{\IP}{\mathbb{P}}%{{\relax{\rm I\kern-.18em P}}}

\newcommand\IZ{\mathbb {Z}}

\newcommand\IQ{\mathbb {Q}}
\newcommand{\IC}{\mathbb{C}}

\newcommand{\IR}{\mathbb{R}}
\newcommand{\CU}{{\mathcal U}}
\newcommand{\ba}{\begin{array}}
\newcommand{\ea}{\end{array}}

\newcommand{\CV}{{\mathcal V}}
\newcommand{\CX}{{\mathcal X}}

\newcommand{\CB}{{\mathcal B}}

\newcommand{\CY}{{\mathcal Y}}
\newcommand{\CW}{{\mathcal W}}

\newcommand{\bal}{\begin{aligned}}
\newcommand{\eal}{\end{aligned}}

\newcommand{\CZ}{{\mathcal Z}}

\newcommand{\longto}{\longrightarrow}

\newcommand{\ch}{{\mathrm{ch}}}

\newcommand{\CO}{{\mathcal O}}

\newcommand{\CA}{{\mathcal A}}
\newcommand{\wCA}{{\widetilde{\mathcal A}}}

\newcommand{\CF}{{\mathcal F}}

\newcommand{\CM}{{\mathcal M}}
\newcommand{\CL}{{\mathcal L}}
\newcommand{\CJ}{{\mathcal J}}
\newcommand{\CR}{{\mathcal R}}
\newcommand{\CC}{{\mathcal C}}
\newcommand{\CI}{{\mathcal I}}

\newcommand{\CQ}{{\mathcal Q}}
\newcommand{\CN}{{\mathcal N}}

\newcommand{\wI}{{\widetilde I}}

\newcommand{\wJ}{{\widetilde J}}

\newcommand{\wZ}{{\widetilde Z}}

\newcommand{\wA}{{\widetilde A}}

\CompileMatrices
\LaTeXdiagrams
\UseAllTwocells
\newdimen\tableauside\tableauside=1.0ex
\newdimen\tableaurule\tableaurule=0.4pt
\newdimen\tableaustep
\def\phantomhrule#1{\hbox{\vbox to0pt{\hrule height\tableaurule width#1\vss}}}
\def\phantomvrule#1{\vbox{\hbox to0pt{\vrule width\tableaurule height#1\hss}}}
\def\sqr{\vbox{%
  \phantomhrule\tableaustep
  \hbox{\phantomvrule\tableaustep\kern\tableaustep\phantomvrule\tableaustep}%
  \hbox{\vbox{\phantomhrule\tableauside}\kern-\tableaurule}}}
\def\squares#1{\hbox{\count0=#1\noindent\loop\sqr
  \advance\count0 by-1 \ifnum\count0>0\repeat}}
\def\tableau#1{\vcenter{\offinterlineskip
  \tableaustep=\tableauside\advance\tableaustep by-\tableaurule
  \kern\normallineskip\hbox
    {\kern\normallineskip\vbox
      {\gettableau#1 0 }%
     \kern\normallineskip\kern\tableaurule}%
  \kern\normallineskip\kern\tableaurule}}
\def\gettableau#1 {\ifnum#1=0\let\next=\null\else
  \squares{#1}\let\next=\gettableau\fi\next}
\title{D-branes, surface operators, and ADHM quiver representations}
\author[U. Bruzzo, W.-y. Chuang, D.-E. Diaconescu, 
M. Jardim, G. Pan,\\ Y. Zhang]{U. Bruzzo${}^{1,2}$, W.-y. Chuang${}^{3}$, 
D.-E. Diaconescu${}^4$, 
M. Jardim${}^5$, G. Pan${}^4$,\\ Y. Zhang${}^4$}
\address{${}^1$ Department of Mathematics, University of Pennsylvania, David Rittenhouse \linebreak
Laboratory, 209 South 33rd Street, Philadelphia, PA 19104, USA${}^\dag$
\footnotetext{${}^\dag$ On leave of absence from International School for Advanced Studies, 
Trieste, Italy} }
\address{${}^2$ 
Istituto Nazionale di Fisica Nucleare, Sezione di Trieste}
\address{${}^3$ Department of Mathematics, National Taiwan University, Taipei, Taiwan}
\address{${}^4$ NHETC, Rutgers University\\
Piscataway, NJ 08854-0849 USA}
\address{${}^5$ IMECC - UNICAMP
Department of Mathematics
Rua S{\'e}rgio Buarque de Holanda, 651
Bar�o Geraldo, Campinas, SP - Brazil
CEP 13083-859}

\begin{document}

\begin{abstract}
A supersymmetric quantum mechanical model is constructed for  
BPS states bound to surface operators in five dimensional $SU(r)$ 
gauge theories using D-brane engineering. 
This model represents the effective action of 
a certain D2-brane configuration, and is naturally obtained by dimensional 
reduction of a quiver $(0,2)$ gauged linear sigma model. 
In a special stability chamber, the resulting moduli space of 
 quiver representations is shown to be  virtually smooth 
and isomorphic to a moduli space of framed quotients on the projective plane. 
A precise conjecture relating a K-theoretic partition function of this moduli 
space to refined open string invariants of toric lagrangian branes is formulated 
for conifold and local $\IP^1\times \IP^1$ geometries. 
\end{abstract}

\maketitle
\tableofcontents

\section{Introduction}
The main goal of this paper is to construct a microscopic quantum mechanical model for BPS states bound to certain surface operators in minimally supersymmetric five dimensional $SU(r)$ gauge theories. This model is obtained employing a string theory construction of such theories 
consisting of 
IIA D-branes in a nontrivial geometric background. The BPS states are engineered in 
terms of D2-brane configurations, the resulting low energy effective action being naturally 
constructed as 
the dimensional reduction of a $(0,2)$ quiver gauged linear sigma model. 
An ADHM style theorem is proven, identifying  
the moduli space of quiver representations 
in a special stability chamber with a moduli space of decorated framed torsion free 
sheaves on the projective plane. The counting function of BPS states bound to surface operators is identified with a K-theoretic partition 
function of this moduli space. 
A precise conjecture is formulated, relating this partition function to 
refined open string invariants of toric lagrangian branes in conifold 
and local $\IP^1\times \IP^1$ geometries. This conjecture is motivated by 
previous work on the subject  \cite{Alday:2009fs,Dimofte:2010tz}, where
surface operators are engineered by branes wrapping such cycles. Previous papers on 
a similar subject also include  \cite{Alday:2010vg,Awata:2010bz,Kozcaz:2010af,Kozcaz:2010yp}, treating various aspects of surface operators in relation with localization on affine Laumon spaces 
and two dimensional conformal field theory. 
The relation between some of these results and the present work will be explained below. 

In more detail, this paper is structured as follows. Five dimensional gauge theories 
are constructed in section (\ref{sectiontwoone}) using D6-branes wrapping exceptional
cycles of a resolved ADE singularity. Surface operators are obtained by adding D4-branes 
wrapping certain supersymmetric cycles in this background.  
BPS states bound to surface operators are identified 
with supersymmetric ground states of a certain D2-brane system with boundary 
on a D4-brane. The effective action of this system is constructed in section 
(\ref{sectiontwotwo}) by dimensional 
reduction of a $(0,2)$ quiver gauged linear sigma model. The final result is given in 
the quiver diagram \eqref{eq:zerotwoquiverC} and the table \eqref{eq:EJfinal}. 

The geometry of the resulting moduli space of flat directions is studied in detail 
in section (\ref{sectionthree}). Theorem (\ref{flatdirthm}) 
proves that the quantum mechanical moduli space is isomorphic to the moduli 
space of $\theta$-stable 
representations of a quiver with relations presented in section (\ref{sectionthreeone}), 
equation \eqref{eq:enhquiverA}. This quiver is an enhancement of the standard ADHM quiver whose stable representations are in one-to-one correspondence to isomorphism classes of framed torsion free sheaves on the projective plane. As opposed to the 
standard ADHM quiver, the space of $\theta$-stability conditions has a nontrivial 
chamber structure. In particular Lemma (\ref{thetachamber}) establishes the existence 
of a special chamber where $\theta$-stability is equivalent with an algebraic stability 
condition generalizing standard ADHM stability.  
Theorem (\ref{smoothmoduli}) proves that the moduli space of stable quiver 
representation is virtually smooth in the special chamber, and provides an explicit 
presentation of its virtual tangent space. Finally, Theorem (\ref{relquotquiver}) 
proves that in the special 
stability chamber the moduli space is isomorphic to a moduli space of data 
$(E,\xi,G,g)$ where $E$ is a torsion free sheaf on the projective plane, 
$\xi:E{\buildrel \sim\over \longto} \CO_{D_\infty}^{\oplus r}$ 
is a framing of $E$ along a hyperplane $D_\infty\subset \IP^2$, and 
$g:E\twoheadrightarrow G$ is a skyscraper quotient of $E$ 
supported (in the scheme theoretic sense) on a fixed hyperplane $D$.  
The fixed hyperplane $D$ represents the support of the surface operator. 
Note that similar moduli spaces (without framing data) 
have been studied by Mochizuki in \cite{parab-hilb,M-inv}. 
The data $(G,g)$ can be also interpreted as a degenerate 
parabolic structure of $E$ along $D$, since only zero dimensional quotients 
of $E|_D$ are involved. In similar situations studied in the literature 
\cite{Alday:2010vg, Awata:2010bz,Kozcaz:2010yp}, surface operators are associated to 
affine Laumon spaces \cite{affine-L}, 
which are moduli spaces of framed parabolic sheaves $E$
on $\IP^1\times \IP^1$. In those cases, the parabolic structure consists of a 
genuine filtration of the restriction $E|_D$, as expected from the general classification
of surface operators \cite{ramification}.
  The moduli space obtained above offers a 
different geometric model for surface operators, its
viability being tested in section (\ref{sectionfive}) by comparison with refined open string 
invariants. The relation between these models will become clearer below, once their connection 
with toric open string invariants is understood. 

The counting function of BPS states is identified 
with a K-theoretic counting function for stable enhanced ADHM quiver representations 
in section (\ref{sectionfour}). The moduli space of stable quiver 
representations is equipped by construction with a natural torus action 
and a determinant line bundle. 
The K-theoretic partition function defined in section (\ref{sectionfourtwo}) 
is a generating 
function for the equivariant virtual Euler characteristic of this determinant line bundle. 

From a physical point of view $(0,q)$-forms on the moduli 
space with values in the determinant line bundle are supersymmetric 
ground states in the quiver quantum mechanics constructed in section 
(\ref{sectiontwo}). 
The torus invariant stable 
quiver representations in the special chamber are classified 
in terms of sequences of nested partitions in 
Proposition (\ref{fixedpointlemma}). Moreover, an explicit expression for the 
equivariant K-theory class of the tangent space at each fixed point is also 
provided. This yields an explicit expression \eqref{eq:equiveulerA} 
for the equivariant virtual  Euler characteristic
 of the determinant line bundle. 

Section (\ref{sectionfive}) consists of a detailed comparison of the 
$r=1,2$ quiver K-theoretic partition functions in the special chamber and 
the refined open string invariants of toric lagrangian branes  
in the corresponding toric threefold $Z$. This relation is stated in Conjecture (\ref{quivrefopenconjA}) for $r=1$, and Conjecture (\ref{quivrefopenconjB}) 
for $r=2$, both conjectures being supported by extensive numerical 
computations. Computation samples are provided in Examples 
(\ref{sampleA}), (\ref{sampleB}).  

Some details of this relation may help elucidate the connection between the present construction and previous work \cite{Dimofte:2010tz, Awata:2010bz}. 
Note that the refined vertex formalism developed in \cite{Iqbal:2007ii} assigns to a special lagrangian cycle $L$ three distinct refined open string partition functions 
corresponding to the choice of a preferred leg of the refined vertex.  
If the brane $L$ is placed on one of the two ordinary legs, the resulting partition functions are related by a simple change of variables. These are cases I and II in 
\cite[Sect 4.2]{Iqbal:2007ii}. In the third case, III,  
 the lagrangian brane is placed on the preferred leg, resulting in a 
different expression for the open topological partition function. The third case has been considered in connection with surface operators in \cite{Dimofte:2010tz}. In particular
the refined topological open string partition function is identified in loc. cit. 
with a surface operator  partition function in the limit 
$\Lambda_{inst}\to 0$. A similar comparison was carried out in  \cite{Awata:2010bz}
for topological, non-refined open string invariants, in which case there is no distinction 
between the three legs. As mentioned above, the surface operator partition function 
is calculated in \cite{Awata:2010bz} by localization on affine Laumon spaces. 

Conjectures (\ref{quivrefopenconjA}) and (\ref{quivrefopenconjB}) establish a 
precise relation between the K-theoretic partition function introduced in section 
(\ref{sectionfour}) and the refined open string partition function of an external 
toric lagrangian brane. This means that the brane intersects only a noncompact 
component of the toric skeleton of the Calabi-Yau threefold $Z$, as discussed in detail 
in section (\ref{sectionfive}). A similar relation is expected between the 
equivariant K-theory partition function of the affine Laumon space and the 
refined open string invariants of an internal toric lagrangian brane 
\cite{Dimofte:2010tz}. An internal brane intersects a compact rational component 
of the toric skeleton of $Z$, therefore such branes are naturally labelled by elements 
of the co-root lattice of the gauge group, in agreement with \cite{ramification}. 
In certain situations, open string invariants of 
external and internal branes can be related by analytic continuation, 
explaining the fact that the same partition function may have different
gauge theoretic constructions. 
 In principle, the surface operators corresponding to internal branes  
can also be engineered as in section (\ref{sectiontwo}), the resulting 
moduli spaces of quiver representations being presumably 
closely related to affine Laumon spaces. This will 
be left for future work. 

{\it Acknowledgements.} We thank Daniel Jafferis and Greg Moore for discussions. We are also grateful to Valeriano Lanza for pointing out an error in the proof of smoothness given in the previous version of the paper. 
UB's work is partially supported by the PRIN project "Geometria delle 
variet\`a algebriche e dei loro spazi di moduli" and INFN project PI14 "Nonperturbative 
dynamics of gauge theories". WYC was partially supported by DOE grant DE-FG02-96ER40959.
The work of DED was partially supported by NSF grant PHY-0854757-2009.
MJ is partially supported by the CNPQ grant number 305464/2007-8 and the
FAPESP grant number 2005/04558-0.

{\it Note Added.} There may be some partial overlap with \cite{W-surface}, which 
appeared when the present work was completed. 

\section{Surface operators and quiver quantum mechanics}\label{sectiontwo}

This section presents a IIA D-brane construction of BPS states in five 
dimensional gauge theories, in the presence of surface operators. 
The final outcome, presented in detail at the 
end of section (\ref{sectiontwotwo}), is a supersymmetric quiver  
quantum mechanical model for such states obtained as the effective 
action of certain D2-brane configurations with boundary.

\subsection{D-brane engineering}\label{sectiontwoone} 
Minimally supersymmetric five dimensional gauge theories can be easily constructed using 
IIA D6-branes wrapping rational holomorphic curves in a K3 surface. More precisely, consider a K3 surface with a canonical ADE singularity. Its crepant resolution contains a configuration of $(-2)$ rational curves whose intersection matrix is determined by the 
incidence matrix of the corresponding Dynkin diagram. A configuration consisting of an arbitrary 
number of D6-branes wrapped on each such curve yields 
in the low energy limit a quiver five dimensional gauge theory with eight supercharges. 
Moreover, BPS states in this quiver gauge theory
can be obtained by wrapping D2-branes on the same holomorphic cycles. 
Then standard D-brane technology shows that the effective action of such a D-brane 
configuration is a supersymmetric quiver quantum mechanics. This is a microscopic model for such BPS states which can be effectively used  in counting problems via localization 
on moduli spaces of stable quiver representations. It will be shown below that a similar
model can be constructed for BPS states bound to a surface operator. 
Since toric geometry methods will be used, only K3 surfaces with $A_{k}$ singularities are amenable to the approach developed below.  Moreover in order to keep the technical details to a minimum, the construction will be carried out only for $k=1$. The same basic principles 
apply to all $k\geq 1$, more involved computations being required. 

For the present purposes it suffices to consider a noncompact K3 surface $T$ 
isomorphic to the total space of the cotangent bundle $T^*\IP^1$. The time direction 
will be Wick rotated to euclidean signature and assumed to be periodic. This yields 
a natural presentation of the BPS counting function as a finite temperature 
partition function. Therefore one obtains a geometric background of the form 
$T\times S^1\times \IR^5$ in IIA theory in euclidean space-time. 
Note that periodic time translations form a free $S^1$-action on the space-time 
manifold. In this setup, the world volume of a D$p$-brane is a 
submanifold of space-time of real dimension $(p+1)$ preserved by the 
free $S^1$ action. In contrast, the world-volume of a D$p$-instanton is a $(p+1)$-submanifold embedded in a fixed time subspace. D$p$-instantons 
will not be employed in the following, therefore all D-brane world-volume manifolds
must be invariant under time translations. 
Let $(x^1,\ldots, x^5)$
be linear coordinates on $\IR^5$. 

Minimally supersymmetric five dimensional $SU(r)$ Yang Mills theory is engineered 
by  $r$ coincident D6-branes with world-volume $\IP^1\times S^1\times \IR^4$, 
where $\IP^1$ is 
identified with the zero section of 
$T\to \IP^1$, and $\IR^4\subset \IR^5$ is a linear subspace.  
Let $(x^1,\ldots, x^5)$ be linear coordinates on $\IR^5$ so that the later is 
the hyperplane $x^5=0$. BPS particles in this theory are engineered by D2-branes with 
world-volume 
$\IP^1\times S^1$. Therefore BPS states are identified to supersymmetric 
ground states in the effective action of D2-branes in the presence of D6-branes, 
which will be explicitly constructed later in this section. 

In order to construct supersymmetric surface operators, note that there is a 
natural identification $T\times S^1\times \IR^5\simeq T\times \IC^\times 
\times \IR^4$, where  $\IR^4\subset \IR^5$ is the hyperplane $x^5=0$. The 
isomorphism $S^1\times \IR\simeq \IC^\times$ is given by 
$U=e^{x^5+i\theta}$, where $\theta$ is an angular coordinate on $S^1$. 
The free $S^1$-action corresponding to euclidean time translations 
is $\theta \to 
\theta +\delta \theta$. Obviously, $T\times \IC^\times$ is a toric Calabi-Yau threefold
preserved by this action. 
Then surface operators will be engineered by wrapping 
D4-branes on $M\times \IR^2$, where $M\subset T\times \IC^\times$ is an $S^1$-invariant 
toric special lagrangian 
and $\IR^2\subset \IR^4$ is the linear subspace $\{x^1=x^2=0\}$. 

The cycle $M$ will be constructed employing the methods used in \cite{mirror-discs}. 
Note that $T$ is a toric quotient $\left(\IC^3\setminus \{X_1=X_2=0\}\right)/\IC^\times$, where $(X_1,\ldots, X_3)$ are linear coordinates on $\IC^3$ 
such that weights of the $\IC^\times$ action are $(1,1,-2)$. Alternatively, $T$ 
admits a presentation as a symplectic quotient $\IC^3//U(1)$ with respect to a hamiltonian $U(1)$ action with moment map 
\[ 
\mu(X_1,\ldots, X_3) = |X_1|^2 +|X_2|^2-2|X_3|^2. 
\]  
The $U(1)$ action on the level set $\mu^{-1}(\zeta)$, $\zeta\in \IR_{>0}$ is 
free and the quotient $\mu^{-1}(\zeta)/U(1)$ is isomorphic to $T$.  Note also that there is a natural symplectic torus action
$U(1)^2\times T\to T$, the resulting moment map giving a projection 
$\varrho:T\to \IR^2$.  The image 
of $\varrho$ is the Delzant polytope of $T$. 
In homogeneous coordinates, this map is given by 
\[
\varrho(X_1,X_2,X_3) =(|X_1|^2,|X_2|^2, |X_3|^2),
\]
where $\IR^2\subset \IR^3$ is identified with the hyperplane
\be\label{eq:momentA}
 |X_1|^2 +|X_2|^2-2|X_3|^2=\zeta.
 \ee
Obviously, there is a similar map 
${\widetilde \varrho} : T\times \IC^\times \to \IR^3$, 
$${\widetilde \varrho}(X_1,X_2,X_3,U)= (|X_1|^2,|X_2|^2, |X_3|^2,|U|^2).$$
Using the methods of 
\cite{mirror-discs}, the cycle $M$ will be constructed by first specifying its image 
under ${\widetilde \varrho}$, 
\be\label{eq:slagA} 
|X_1|^2 -|X_2|^2 = c_1, \qquad 
|U|^2 - |X_2|^2 = c_2,
\ee
 where $c_1,c_2$ are real parameters. Suppose 
 \[ 
 c_1>\zeta > 0, \qquad c_2>0. 
\]
Then, taking into account equation \eqref{eq:momentA}, it follows that 
any solution to \eqref{eq:slagA} must satisfy the inequalities 
\[
|X_1|^2 \geq c_1, \qquad |X_2|^2 \geq 0, \qquad |X_3|^2 \geq {1\over 2}(c_1-\zeta),\qquad |U|^2 \geq c_2.
\]
Therefore the image of $M$ under ${\widetilde \varrho}$ is a half real line. 
and $X_1, X_3$ are not allowed to 
vanish for any solution to \eqref{eq:slagA}. 
$M$ is defined by specifying linear constraints on the phases of the homogeneous 
coordinates in addition to equations \eqref{eq:slagA}. 
 The intersection of $M$ with the dense open subset 
$X_2\neq 0$ is a union of two two-tori defined by the equations 
\be\label{eq:slagB} 
\phi_1+\phi_2+\phi_U=0,\pi.
\ee
The intersection of $M$ 
 with the divisor $X_2=0$ is the two-torus 
 \be\label{eq:slagC} 
 |X_1|^2 =c_1,  \qquad |X_3|^2 ={1\over 2}(c_1-\zeta),\qquad |U|^2 = c_2,
 \ee
 the phases of $X_3,U$ being unconstrained, while the phase of $X_1$ 
 is set to zero using $U(1)$ gauge transformation. Therefore the two branches 
 of $M$ defined in equation \eqref{eq:slagB} are joined together 
 at $X_2=0$, resulting in a special lagrangian cycle of the form $T^2\times \IR$. 
 Taking a single branch would yield a special lagrangian cycle with boundary, 
 $T^2\times \IR_{\geq 0}$.
 
  For further reference  note that there is a one parameter family of 
  holomorphic discs in 
 $T\times \IC^\times$ with boundary on $M$ cut by the equations 
 \be\label{eq:holdisc}
 X_2=0, \qquad 0\leq |X_3| \leq {1\over 2}(c_1-r), \qquad U = \sqrt{c_2}e^{i\theta}. 
 \ee
 Note also that $M$ is invariant under euclidean time translations, 
 $\phi_U \to \phi_U + \delta \phi_U$, since any such translation is compensated by a 
 $U(1)$-gauge transformation $\phi_1\to \phi_1-\delta \phi_U$ in 
 \eqref{eq:slagB}. 
 The same $S^1$-action acts freely and transitively on the total space of  
 the family of discs \eqref{eq:holdisc},  identifying the parameter space of this 
 family with the euclidean time circle. 
   
Returning to gauge theory, surface operators are engineered by 
a D4-brane with world-volume $M\times \{x^1=x^2=0\}$. BPS particles bound to
this operator are D2-brane configurations consisting of $n_1$ D2-branes with 
world-volume 
\be\label{eq:branesA} 
x^1=\cdots = x^4=0,\qquad X_3=0, \qquad |U|=\sqrt{c_2}
\ee
and $n_2$ D2-branes with world-volume 
\be\label{eq:branesB} 
x^1=\cdots =x^4=0,\qquad  X_2=0, \qquad 0\leq |X_3| \leq {1\over 2}(c_1-r), \qquad U = \sqrt{c_2}.
\ee
 These stacks of $D_2$-branes will be denoted by D$2_1$, 
 D$2_2$ respectively. Note that the three cycles \eqref{eq:branesA}, \eqref{eq:branesB} 
 are preserved by euclidean time translations, as expected. Taking quotient by this 
 free action yields  in the first case the two-cycle 
 \be\label{eq:branesC}
x^1=\cdots = x^4=0,\qquad X_3=0, \qquad x^5=\mathrm{ln} \, \sqrt{c_2}
\ee
which is isomorphic to the zero section of $T\to \IP^1$. 
In the second case, one obtains  a holomorphic disc $\Delta\subset T$  cut by the equations 
\be\label{eq:branesD}
x^1=\cdots =x^4=0,\qquad  X_2=0, \qquad 0\leq |X_3| \leq {1\over 2}(c_1-r), \qquad 
x^5=\mathrm{ln} \, \sqrt{c_2}.
\ee
This is obviously a vertical  holomorphic disc embedded in the fiber of $T\to\IP^1$ at $X_2=0$. 
Therefore the first stack of D2-branes is wrapped on the zero section 
of $\IP^1$, while the second stack is wrapped on the disc $\Delta$.

\subsection{D2-brane effective action 
via quiver $(0,2)$ models}\label{sectiontwotwo}
To summarize the construction in the previous section, five dimensional 
supersymmetric $SU(r)$ gauge theory is engineered by wrapping $r$ D6-branes 
on the exceptional cycle of a resolved $A_1$ singularity $T$. The space-time is 
Wick rotated to euclidean signature, and the time direction is periodic. 
Surface operators in this theory are engineered by certain supersymmetric 
D4-brane configurations determined by equations \eqref{eq:slagA}, 
\eqref{eq:slagB}. BPS states bound to such operators are  
realized by two stacks of  D2-branes 
with multiplicities $n_1,n_2$ wrapping the  holomorphic cycles 
\eqref{eq:branesC}, \eqref{eq:branesD}, which intersect transversely at the 
point $X_2=X_3=0$ in $T$. 

The goal of the present section is to construct the effective action of the stacks 
of D2-branes in this background, including modes of D2-D4 and D2-D6 open strings. 
 Since the D2-branes wrap 
compact cycles, KK reduction will yield an effective quantum mechanical action for 
their zero modes. 
In order to analyze the dynamics of this D-brane system, 
it is helpful to note that that it is related
 to the D0-D4-D8-brane configuration studied in 
\cite{gaugefields}. The 
effective action of the D0-branes was identified in  \cite{gaugefields} with
 a gauged version of the
$(0,4)$ ADHM sigma model action constructed in  \cite{adhmsigma}.

As opposed to the current case, the D-brane system analyzed in \cite{gaugefields} 
is embedded in flat space. In order to understand the relation between these configurations, 
the complex surface $T\to \IP^1$ must be replaced by $T'=T^2\times \IC\to \IC$, allowing 
two flat space directions to  be compact. 
Then consider a D$2_1$-D$2_2$-D6-brane system in the new background 
 consisting of $n_2$ D2-branes wrapping a $T^2$ fiber of $T'\to \IC$, 
 $n_1$ D2-branes wrapping a section of $T'\to \IC$, and $r$ D6-brane wrapping the 
 same section and a linear subspace $\IR^{4}\subset \IR^{5}$.  Obviously, the 
 relative positions of these branes are the same as the relative positions in the 
 D$2_1$-D$2_2$-D6 system on $T$. 
  The new brane system on $T'\times \IR^{5}$ 
is related by a T-duality transformation on $T^2$ 
to the configuration of parallel D0-D4-D8 branes studied in \cite[Sect 3]{gaugefields}.
The D0-brane effective action was constructed there
by dimensional reduction of a two dimensional 
$(0,4)$ gauged linear sigma model, obtaining a quantum mechanical action 
with four supercharges. In the present case, $T'$ is replaced by $T$, which breaks half 
of the underlying thirty-two IIA supercharges, and in addition a D4-brane is added to the system. The resulting configuration preserves only two supercharges as opposed to four. 
Therefore by analogy with \cite{gaugefields}, 
 the effective action  will be 
 constructed by dimensional reduction of a two dimensional 
 $(0,2)$ gauged linear sigma model
  \cite[Sect. 6]{twodphases}. 
  Since the system is fairly complicated, it will be convenient to proceed in several stages. 
  The D$2_1$-D6, D$2_2$-D4 configurations will be first studied separately, classifying the 
  massless states in $(0,2)$-multiplets (reduced to one dimension), and writing down the 
  interactions in $(0,2)$ formalism. The coupling between these two sectors via 
  open string D$2_1$-D$2_2$ massless modes will be studied at the next stage. 
  In the following all Chan-Paton bundles on branes will be taken topologically  trivial.

  \subsubsection{$(0,2)$ models}\label{twotwoone}
  Since the massless states will be classified in $(0,2)$ multiplets reduced to 
  one dimension, a brief review of such models is provided below, following 
\cite[Sect 6.1]{twodphases}. There are three types of $(0,2)$ multiplets, 
the chiral multiplet, the Fermi multiplet, and the vector multiplet. 
The on shell $(0,2)$ chiral multiplet 
consists of a complex scalar field and a complex chiral fermion of positive chirality, while 
the $(0,2)$ Fermi multiplet consists of a complex chiral fermion of negative chirality. 
The  $(0,2)$ gauge multiplet consists of a gauge field and an adjoint  complex chiral fermion. 
A pair consisting of one $(0,2)$-chiral multiplet and one $(0,2)$ Fermi multiplet 
has the same degrees of freedom as  a $(2,2)$ multiplet \cite[Sect 6.1]{twodphases}. 
Chiral multiplets will be denoted by $\CA_+$ in the following, and Fermi multiplets 
will be denoted by $\CY_-$. 
Each Fermi superfield $\CY_-$ satisfies a superspace constraint of the form 
\be\label{eq:susyconstrA}
{\overline {\mathcal D}_+} {\mathcal Y}_- = \sqrt{2}E_{\CY_-}, 
\qquad {\overline {\mathcal D}_+}E_{\CY_-}=0, 
\ee
where $E_{\CY_-}$ is a holomorphic function of chiral superfields taking values in the 
same representation of the gauge group as $\CY_-$. 
Additional F-term like interactions can be written down in terms of 
some holomorphic functions $J_{\CY_-}$ of chiral superfields which take values 
in the dual representation of the gauge group.  The following constraint 
\be\label{eq:susyconstrB} 
\sum_{\CY_-} \langle J_{\CY_-} \, ,\, E_{\CY_-} \rangle =0.
\ee
must be satisfied in order to obtain a $(0,2)$ supersymmetric lagrangian.
Then the $(0,2)$ superspace action is \cite[Sect 6.1]{twodphases}
\be\label{eq:susyactA} 
\bal 
& {1\over 8} \int d^2x d\theta^+d\theta^+ \mathrm{Tr}(\CW\CW) 
-{i\over 2} \int d^2x d^2\theta \sum_{\CA} {\overline \CA}
({\mathcal D}_0 - {\mathcal D}_1) \CA \\
& 
-{1\over 2} \int d^2x d^2\theta \sum_{\CY_-}\CY_-^\dagger \CY_-
-{1\over \sqrt{2}} \int d^2x d\theta^+ \sum_{\CY_-} 
\langle J_{\CY_-} \, ,\, {\CY_-} \rangle |_{{\overline \theta}^+},\\
\eal 
\ee
where $\CW$ is the field strength of the vector multiplet. 
In addition, one can add an FI term of the form 
\[
{\zeta\over 4} \int d^2x d\theta^+ \mathrm{Tr}\CW|_{{\overline {\theta^+}}=0} 
+ h.c
\]
for each simple factor of the gauge group.
The total potential energy of the resulting $(0,2)$ lagrangian is 
\be\label{eq:potentialA} 
U_D + \sum_{\CY_-} |E_{\CY_-}|^2 + |J_{\CY_-}|^2 
\ee
where $U_D$ is a standard D-term contribution.  Moreover, assuming that $E_{\CY_-}, 
J_{\CY_-}$ are polynomial functions in the chiral superfields $\CA$, the Yukawa couplings 
can be written as follows. Any monomial $\CA_{1}\cdots \CA_n$
in $E_{\CY_-}$ determines a sequence of Yukawa couplings of the form 
\be\label{eq:yukawaAi} 
\sum_{i=1}^n \langle \lambda_{\CY_-}^\dagger, A_1\cdots A_{i-1}\psi_{\CA_i} \CA_{i+1} 
\cdots A_n \rangle
\ee
and any monomial $\CA_{1}\cdots \CA_n$
in $J_{\CY_-}$ determines a sequence of Yukawa couplings of the form 
\be\label{eq:yukawaAii} 
\sum_{i=1}^n \langle \lambda_{\CY_-}, A_1\cdots A_{i-1}\psi_{\CA_i} \CA_{i+1} 
\cdots A_n\rangle.
\ee
Then one has to sum over all $\CY_-$ and over all such monomials.

 \subsubsection{D$2_1$-D6 system}\label{twotwotwo}
 Recall that this D-brane system is supported on the zero section of $T$, and the
 Chan-Paton bundles $E_1,F$ are topologically trivial. In addition,  
 $E_1,F$ are equipped with hermitian structures and compatible connections, which 
 determine in particular holomorphic structures.  Since they are bundles on $\IP^1$, 
 $E_1,F$ must be isomorphic to the trivial holomorphic bundles $V_1\otimes \CO_{\IP^1}, 
 W\otimes \CO_{\IP^1}$, where $V_1,W$ are vector spaces of dimensions $n_1,r$
 equipped with hermitian structures. Moreover, the Chan-Paton connections are gauge equivalent to the trivial connection. The temporal component of the gauge field has a 
  constant zero mode on $\IP^1$. 
  
  The normal bundle to the D2-branes in $T\times \IR^5$ is 
 $N_1\simeq \CO_{\IP^1}(-2)
 \oplus \CO_{\IP^1}^{\oplus 2} \oplus \CR_{\IP^1}$, where $\CR_{\IP^1}$ 
 denotes the trivial real line bundle. 
 The transverse fluctuations of the D2-brane are parametrized by a section 
 $(\Phi_1,A_1,A_2,\sigma_1)$ of $N_1\otimes {End}(E_1)$, the last 
 component, $\sigma_1$, being subject to the reality condition
 $\sigma_1^\dagger =\sigma_1$. 
 Then the zero modes of the transverse fluctuations  are holomorphic sections of 
 $\mathrm{End}(V_1)\otimes \left(\CO_{\IP^1}(-2)
 \oplus \CO_{\IP^1}^{\oplus 2}\right)$. Therefore $\Phi_1$ is identically zero,
 and $A_1,A_2,\sigma_1$ are constant. 
 
 In conclusion KK reduction on $\IP^1$ 
 yields two  complex fields $A_1,A_2\in \mathrm{End}(V_1)$, a real field 
 $\sigma_1 \in \mathrm{End}(V_1)$, and an $U(V_1)$-gauge field. 
  This is precisely the bosonic field content of an $D=4$, $N=2$
   vector multiplet reduced to one dimension, which is 
   expected since the D2-branes preserve eight supercharges. By supersymmetry 
   the fermionic fields are obtained by dimensional reduction of the fermions in the same multiplet. The resulting massless spectrum can be organized in terms of two dimensional $(0,2)$ multiplets reduced to one dimensions. Namely, there are two complex adjoint 
   $(0,2)$ chiral superfields, $\CA_{1+},\CA_{2+}$ with bosonic components $A_1,A_2$, 
   a $(0,2)$ vector multiplet, and a $(0,2)$ adjoint Fermi multiplet $\CX_-$.
   The gauge fields and real adjoint bosonic field $\sigma_1$ are obtained by reduction of 
   the two dimensional vector multiplet. 
   
A similar analysis must be carried out for the D$2_1$-D6 fields. In flat space space, 
with trivial Chan-Paton bundles, and trivial gauge connections, 
the massless open string modes in this sector yield a $D=3$, $N=4$ bifundamental 
hypermultiplet on the D2-brane world-volume. There are 
two complex bosonic fields $I,J$, sections of $Hom(F,E_1)$, 
$Hom(E_1,F)$ respectively,  and two bifundamental Dirac fermions $\psi, {\widetilde \psi}$, 
also sections of $Hom(F,E_1)$, 
$Hom(E_1,F)$. Note that there is an $SU(2)_R$ global symmetry group induced by 
transverse rotations to the D2-D6 system. 
The bosonic fields are $SU(2)_R$-singlets, while the fermions $(\psi, {\widetilde \psi}^\dagger)$ form a doublet. 
When the D-branes are wrapped on the zero section of $T\to \IP^1$, 
the bosonic fields $I,J$ are still sections $I,J$ of 
${Hom}(F,E_1)$, ${Hom}(E_1,F)$, which have constant zero modes 
on $\IP^1$. Therefore KK reduction on $\IP^1$  
yields two complex bosonic fields $I,J$ 
with values in $\mathrm{Hom}(W,V_1)$, $\mathrm{Hom}(V_1,W)$ respectively.

The fermions are topologically twisted as follows.  
The Lorentz symmetry group $Spin(3)\simeq SU(2)$ and the global symmetry group 
$SU(2)$ are broken to $U(1)$ subgroups identified with the spin groups of 
the tangent, respectively normal bundle to the zero section. Both fermions $\psi$, 
${\widetilde \psi}$ 
have $U(1)\times U(1)$ charges $(1,1)\oplus (-1,1)$. Moreover, the normal bundle 
is canonically identified with the cotangent bundle 
of the zero section once a global holomorphic 2-form on $T$ is chosen. 
Since the cotangent bundle is dual to the tangent bundle, it follows that the components
of ${\psi}, {\widetilde \psi}$ are sections of  
$${Hom}(F,E_1)\otimes \left(\CO_{{\IP^1}}\oplus \CO_{\IP^1}(-2)\right),\qquad 
  {Hom}(E_1,F)\otimes \left(\CO_{{\IP^1}}\oplus \CO_{\IP^1}(-2)\right)$$
   respectively. Therefore dimensional reduction on $\IP^1$ yields two chiral fermion fields 
    with values in $Hom(W,V_1)$, $Hom(V_1,W)$, which are related by supersymmetry 
    to the bosonic fields. In conclusion, the D$2_1$-D6 strings yield two 
    $(0,2)$ chiral superfields $\CI_+,\CJ_+$ with values in $Hom(W,V_1)$, $Hom(V_1,W)$
    and no other degrees of freedom. 

Finally, it is helpful to note that there is an alternative derivation of the D$2_1$-D6 massless 
spectrum, following from the observation that $T$ is the crepant resolution of the $\IC^2/\IZ_2$ orbifold 
singularity. Then D-branes wrapped on the zero section with trivial Chan-Paton bundles 
are identified with orbifold fractional branes \cite{DM,fractbr} associated to the trivial 
representation of the orbifold group.  
More specifically, the D$2_1$-branes are identified with $n_1$ fractional D0-branes, 
while the D6-branes are identified with $r$ fractional D4-branes. 
Therefore the massless open string spectrum is identified with the 
$\IZ_2$-invariant part of the spectrum of a D0-D4 system transverse to the orbifold, 
the action of orbifold group on the Chan-Paton 
spaces $V_1$, $W$ being trivial. Then a straightforward computation similar to 
\cite{DM} yields the same massless spectrum as obtained above by geometric 
methods. In particular, all transverse fluctuations of the D0-branes
along the orbifold directions are projected out. 
The field content of the effective action is encoded in the following 
quiver diagram 
\be\label{eq:zerotwoquiverA}
\xymatrix{
 V_1 
  \ar@(ur,ul)_-{\CA_{1+}}\ar@(dl,dr)_-{\CA_{2+}} \ar@(ul,dl)_-{\CX_-}
  \ar@/^/[rr]^-{\CI_+}
&&  W. \ar@/^/[ll]^-{\CJ_+}\\}
\ee
where each arrow represents a $(0,2)$ multiplet reduced to one dimension. 

As explained in section $(2.2.1)$, the interactions are determined by two 
holomorphic functions $E_{\CX_-}$, $J_{\CX_-}$ of the chiral superfields 
$\CA_{1+}, \CA_{2+}, \CI_+, \CJ_+$. Their tree level values can be easily determined 
using the fractional brane description of the system explained in the previous 
paragraph. The tree level potential energy of the D$2_1$-D6 system is the same as the 
tree level potential energy of a flat space D0-D4 system, truncated to $\IZ_2$-invariant 
fields. This yields the following expression 
\be\label{eq:FDpotential} 
|[A_1,A_2]+IJ|^2+
|[A_1,A_1^\dagger ] +[A_2,A_2^\dagger]+II^\dagger-J^\dagger J -\zeta_1|^2, 
\ee
which consists of standard F-term, respectively  D-term contributions. 
$\zeta_1$ is an FI parameter which can be identified with a flat B-field 
background on the D4-brane world-volume. 
The F-term contribution to \eqref{eq:FDpotential} determines 
\be\label{eq:EJb} 
J_{\CX_{-}} =   [\CA_{1+},\CA_{2+}]+\CI_+\CJ_+ , \qquad 
E_{\CX_-} = 0
\ee
up to an ambiguity exchanging $E_{\CX_-}$ and $J_{\CX_-}$. 
In the present context, exchanging $E_{\CX_-}$ and $J_{\CX_-}$ is equivalent to a 
field redefinition, hence there is no loss of generality in making the choice \eqref{eq:EJb}. 
One can also multiply $J_{\CX_-}$ by an arbitrary phase, but this ambiguity can be 
again absorbed by a field redefinition. 

\subsubsection{D$2_2$-D4 system}
By construction, the D4-brane world-volume is of the form $M\times \IR^2$, where 
$M\subset T\times  S^1\times \IR \simeq T\times \IC^\times$ is the $S^1$-invariant 
special 
lagrangian cycle constructed in \eqref{eq:slagA}-\eqref{eq:slagB}. 
The world-volume of the second stack of the D2-branes is the family of 
holomorphic discs \eqref{eq:holdisc} parameterized by periodic euclidean time.  
For fixed time, the D2-branes wrap the 
vertical holomorphic disc in $\Delta \subset T$ given in \eqref{eq:branesD}. 
The geometric background $T\times S^1\times \IR^5$ preserves half of the 
thirty-two IIA supercharges, and the D4-brane wrapped on $M$ preserves only four. 
The combined D$2_2$-D4 system preserves half of the remaining four supercharges. 

The D2-brane fluctuations consist of the standard gauge field, transverse fluctuations, 
and their superpartners. The Chan-Paton bundle $E_2$ is again topologically trivial, 
therefore it can be taken of the form $E_2=V_2\otimes \CO_\Delta$, with 
$V_2$ an $n_2$-dimensional vector space equipped with hermitian structure. 
The Chan-Paton connection is gauge equivalent to the trivial   connection. 
The temporal component of the gauge field has again a constant zero mode on $\Delta$. 

The normal bundle to $\Delta \subset T\times \IR^5$ is trivial, 
$$N_2\simeq \CO_{\Delta} \oplus \CO_{\Delta}^{\oplus 2} \oplus \CR_\Delta,$$
where the first summand is the normal bundle to $\Delta$ in $T$. The second 
and third summands  correspond to the remaining 
five transverse directions, $\CR_\Delta$ denoting the trivial real line bundle 
on $\Delta$.  The transverse fluctuations 
are parameterized therefore by a section $(\Phi_2,B_1,B_2,\sigma_2)$ of 
$\mathrm{End}(V_2)\otimes N_2$, the last component being real, $\sigma_2=\sigma_2^\dagger$. 

In order to determine the zero modes of the transverse fluctuations, boundary 
conditions must be specified for the fields $(\Phi_2,B_1,B_2,\sigma_2)$. 
The fluctuations $\Phi_2,B_1$ are transverse to the D4-brane world-volume, 
therefore they have to satisfy Dirichlet boundary conditions, $\Phi_1|_{\partial \Delta}=0$, 
$B_1|_{\partial \Delta}=0$. This implies that they have no zero modes on $\Delta$ since any holomorphic function which vanishes on the boundary must vanish everywhere. 
The remaining fluctuations $B_2,\sigma_2$ are parallel to the D4-brane, therefore they have to satisfy Newmann boundary conditions. A holomorphic function on $\Delta$ satisfying 
Newmann boundary conditions must be constant, therefore $B_2,\sigma_2$ have constant zero modes on $\Delta$. 

In conclusion, KK reduction on the disc yields a spectrum of bosonic fields consisting of a 
complex field $B_2\in \mathrm{End}(V_2)$, a real field $\sigma_2\in \mathrm{End}(V_2)$ 
and a $U(V_2)$-gauge field. These are the bosonic components of a $(0,2)$ chiral 
multiplet $\CB_{2+}$, and a $(0,2)$ vector multiplet, reduced to one dimension. 
Since the system D$2_2$-D4 preserves two supercharges, the zero modes of the 
fermionic fields must naturally provide the missing fermionic components in these 
multiplets. The resulting field content is summarized in the following quiver diagram
\be\label{eq:zerotwoquiverB} 
\xymatrix{
 V_2 
  \ar@(ul,dl)_-{\CB_{2+}}
\\}
\ee
Since there are no Fermi superfields, the only interactions are gauge 
couplings and D-term interactions. 
This is consistent with the fact that the D2-branes are free to glide along the 
D4 branes with no cost in energy. Note that FI terms in the D2-brane world-volume 
can be obtained by turning on a flat gauge field background on the D4-brane. 

\subsubsection{Coupling the two systems} 
The next task is to couple the two D-brane systems analyzed above. In addition to the 
zero modes found in sections (2.2.2), (2.2.3), there are extra massless open string 
states  in the  D$2_1$-D$2_2$ sector and in the D$2_2$-D6-sector. 
In both cases the stacks of D-branes intersect transversely at a point, therefore the 
massless states are the same as in a similar D-brane configuration embedded in flat space. 
The fields in the D$2_1$-D$2_2$ sector are naturally identified with the components of a 
$D=4$, $N=2$ bifundamental hypermultiplet reduced to one dimension. In terms of $(0,2)$-superfields,  there are two $(0,2)$ chiral multiplets $\Phi_+,\Gamma_+$ 
with values  in $\mathrm{Hom}(V_2,V_1)$, $\mathrm{Hom}(V_1,V_2)$ respectively, 
and two Fermi superfields $\Omega_-,\Psi_-$, also with values in in $\mathrm{Hom}(V_2,V_1)$, $\mathrm{Hom}(V_1,V_2)$. 
The the D$2_2$-D6-sector yields a single Fermi superfield $\Lambda_-$ with values in 
$\mathrm{Hom}(V_2,W)$. Taking into account the previous results, the combined  
$(0,2)$ spectrum is summarized in the following quiver diagram 
\be\label{eq:zerotwoquiverC}
\xymatrix{
 \ar@(ul,dl)_-{\CB_{2+}} V_2 \ar@/^/[rr]^-{\Phi_+, \Omega_-} 
 \ar@/^4pc/[rrrr]^-{\Lambda_-}
 & & V_1 
 \ar@/^/[ll]^-{\Gamma_+, \Psi_-}
  \ar@(ur,ul)_-{\CA_{1+},\CA_{2+}}\ar@(dl,dr)_-{\CX_-} 
  \ar@/^/[rr]^-{\CI}
&&  W. \ar@/^/[ll]^-{\CJ}\\}
\ee
Note that an arrow marked by two superfields represents in fact two distinct arrows, 
corresponding respectively to the two superfields. For ease of exposition, the arrows 
corresponding to chiral superfields will be called bosonic, while the arrows corresponding 
to Fermi superfields will be called fermionic. Therefore, for example, there are three 
arrows beginning and ending at $V_1$, two bosonic corresponding to 
$\CA_{1+}, \CA_{2_+}$, and one 
fermionic, corresponding to $\CX_-$. Similarly, there are two arrows between $V_2$ and $V_1$, one bosonic and one fermionic, and two arrows between $V_1$ and $V_2$, 
again, one bosonic and one fermionic. 

Next one has to determine the holomorphic functions $E,J$ for each Fermi superfield 
in \eqref{eq:zerotwoquiverC}. First note that the tree level potential 
energy must include quartic couplings between the fields $\Phi_+,\Gamma_+$  
superfields $\CA_{1+}, \CA_{2+}, \CB_2$ reflecting the fact that the D$2_1$-D$2_2$ 
fields become massive once the two stacks of D2-branes are displaced, their mass 
being proportional with the separation. 
Therefore, taking into account gauge invariance, the potential interactions between the bosonic components of 
$\Phi_+,\Gamma_+$ and 
$\CA_{1+}, \CA_{2+}, \CB_2$ must be of the form 
\be\label{eq:potentialA} 
|A_1f|^2 + |A_2f-fB_2|^2 + |gA_1|^2 +|gA_2-B_2g|^2.
\ee
Here $f\in \mathrm{Hom}(V_2,V_1)$, $g\in \mathrm{Hom}(V_1,V_2)$ are 
the bosonic components of chiral superfields $\Phi_+,\Gamma_+$. 
Note that since $V_1,V_2,W$ are equipped with hermitian structures, 
any space of morphisms between any two vector spaces has an induced 
hermitian structure. The resulting hermitian form is denoted by $|{}\ \ |$ in  \eqref{eq:potentialA}. 
Such couplings are obtained by setting 
\be\label{eq:EJc} 
\bal 
 E_{\Omega_-} = \epsilon_1(\Phi_+\CB_{2+}-\CA_{2+}\Phi_+), 
 \qquad & J_{\Omega_-} =  \eta_1\Gamma_+\CA_{1+},\\ 
E_{\Psi_-} =\epsilon_2(\CB_{2+}\Gamma_+ - \Gamma_+\CA_{2+}), 
\qquad & J_{\Psi_-} = \eta_2 \CA_{1+}\Phi_+\\
\eal 
\ee
where $\epsilon_1,\epsilon_2,\eta_1,\eta_2$ are phases, i.e. 
complex numbers with modulus 1. One can also obtain the same potential 
energy exchanging the ordered pairs $(E_{\Omega_-}, J_{\Omega_-})$, 
$(J_{\Psi_-}, E_{\Psi_-})$. This ambiguity is equivalent to a field redefinition, 
hence there is no loss of generality in making the choice \eqref{eq:EJc}. 

The phases will be fixed up to field redefinitions imposing the supersymmetry condition 
\eqref{eq:susyconstrB}. Since the coupling between the two sectors will not change the tree level potential 
energy \eqref{eq:FDpotential} of the D$2_1$-D6 modes, 
one must have 
\be\label{eq:EJe} 
J_{\CX_-}= [\CA_{1+},\CA_{2+}]+\CI_+\CJ_+
\ee
as found in equation \eqref{eq:EJb}. The holomorphic function $E_{\CX_-}$ is not necessarily 
zero, as found there, but, if nonzero, it must have nontrivial dependence on the extra chiral superfields $\Phi_+, \Gamma_+$. 

The supersymmetry condition \eqref{eq:susyconstrB} yields  
\be\label{eq:susyconstrD} 
\bal 
\langle J_{\Omega_-}, E_{\Omega_-}\rangle + 
\langle J_{\Psi_-}, E_{\Psi_-}\rangle  + 
\langle J_{\CX_-}, E_{\CX_-}\rangle + 
\langle J_{\Lambda_-}, E_{\Lambda_-}\rangle  =0.
\eal
\ee 
The possible contributions to the holomorphic functions 
$E_{\CY_-}, J_{\CY_-}$ assigned to each Fermi superfield 
$\CY_- \in \{\CX_-,\Omega_-,\Psi_-,\Lambda_-\}$ 
can be classified as follows.  
Let $V_{t(\CY_-)}$, $V_{h(\CY_-)}$ be the vector spaces assigned to the 
tail, respectively the head of the arrow corresponding to $\CY_-$ in the diagram 
\eqref{eq:zerotwoquiverC}. Then $\CY_-$ takes values in the linear space 
$\mathrm{Hom}(V_{t(\CY_-)}, V_{h(\CY_-)})$. The holomorphic functions 
$$E_{\CY_-}\in 
\mathrm{Hom}(V_{t(\CY_-)}, V_{h(\CY_-)}),\qquad J_{\CY_-}\in 
\mathrm{Hom}(V_{h(\CY_-)}, V_{t(\CY_-)})$$
 are determined by linear combinations of paths of bosonic arrows
 in the path algebra of the quiver 
\eqref{eq:zerotwoquiverC}. 

Next note that a simple computation yields 
\be\label{eq:susyconstrC} 
\bal & \langle J_{\Omega_-}, E_{\Omega_-}\rangle + 
\langle J_{\Psi_-}, E_{\Psi_-}\rangle  = 
 (\epsilon_1\eta_1+\epsilon_2\eta_2) 
 \mathrm{Tr}_{V_2} \big(\Gamma_+\CA_{1+}\Phi_+\CB_{2+}\big)\\ &  -\epsilon_1\eta_1
 \mathrm{Tr}_{V_2}\big(\Gamma_+\CA_{1+}\CA_{2+}\Phi_+\big)
 -\epsilon_2\eta_2 \mathrm{Tr}_{V_2}\big(\Gamma_+\CA_{2+}\CA_{1+}\Phi_+\big).\\
 \eal 
\ee
Moreover 
\[
\langle J_{\CX_-}, E_{\CX_-}\rangle = 
\mathrm{Tr}_{V_1}\Big(( [\CA_{1+},\CA_{2+}]+\CI_+\CJ_+) E_{\CX_-}\Big)
\]
where $E_{\CX_-}$ must be a linear combination of paths consisting of the following building blocks 
\[
\Phi_+\CB_{2+}^k\Gamma_+, \qquad \CA_{1+},\qquad \CA_{2+},\qquad  \CI_+\CJ_+,
\]
with $k\in \IZ_{\geq 0}$. 
Similarly, $E_{\Lambda_-}$, $J_{\Lambda_-}$ must be linear combinations 
of paths of the form 
\[
\bal
&\CB_{2+}^{k}\Gamma_+P(\CA_{1+},\CA_{2+}, \CI_+\CJ_+, \Phi_+\Gamma_+,\Gamma_+\Phi_+)\CI_+, \\
&
\CJ_+Q(\CA_{1+},\CA_{2+},\CI_+\CJ_+,\Phi_+\Gamma_+,\Gamma_+\Phi_+)\Phi_+ 
\CB_{2+}^{l}, \\
\eal
\]
where $k,l\in \IZ_{\geq 0}$ and $P(\CA_{1+},\CA_{2+},\CI_+\CJ_+)$,    
$Q(\CA_{1+},\CA_{2+}, \CI_+\CJ_+)$
are polynomial functions of $\CA_{1+},\CA_{2+}, \CI_+\CJ_+, \Phi_+\Gamma_+,\Gamma_+\Phi_+$.
This implies that  
\be\label{eq:susyconstrE}
\langle J_{\CX_-}, E_{\CX_-}\rangle + 
\langle J_{\Lambda_-}, E_{\Lambda_-}\rangle
\ee
cannot contain any terms proportional to 
\[
\mathrm{Tr}_{V_2} \big(\Gamma_+\CA_{1+}\Phi_+\CB_{2+}\big)=
\mathrm{Tr}_{V_1} \big(\Phi_+\CB_{2+}\Gamma_+\CA_{1+}\big).
\]
Therefore supersymmetry requires $\epsilon_1\eta_1+\epsilon_2\eta_2=0$ 
in \eqref{eq:susyconstrC}. 
Then the  remaining terms in the right hand side of 
\eqref{eq:susyconstrC} can be written as 
\[ 
\epsilon_2\eta_2 \mathrm{Tr}_{V_1}\big([\CA_{1+},\CA_{2+}]\Phi_+\Gamma_+\big).
\]
These terms 
must be cancelled by similar terms in the expansion 
of \eqref{eq:susyconstrE}. Since all terms in the expansion of 
$\langle J_{\Lambda_-}, E_{\Lambda_-}\rangle$ have non-trivial dependence 
on $\CI_+, \CJ_+$, the  terms 
required by this cancellation 
must occur in the expansion of $\langle J_{\CX_-}, E_{\CX_-}\rangle$. 
This uniquely determines 
\be\label{eq:EJf} 
E_{\CX_-} = -\epsilon_2\eta_2 \Phi_+\Gamma_+.
\ee
Taking into account all conditions obtained so far, the right hand side of 
\eqref{eq:susyconstrC} reduces to 
\[ 
\langle J_{\Lambda_-}, E_{\Lambda_-}\rangle - \epsilon_2\eta_2 
\mathrm{Tr}_{V_2}( \Gamma_{+}\CI_+\CJ_+ \Phi_+).
\]
Given the building blocks for $E_{\Lambda_-}$, $J_{\Lambda_-}$ listed above, 
it follows that 
\be\label{eq:EJg} 
E_{\Lambda_-} = \epsilon_3 \CJ_+\Phi_+,\qquad 
J_{\Lambda_-} = \eta_3 \Gamma_+\CI_+
\ee
where $\epsilon_3,\eta_3$ are phases satisfying 
$\epsilon_3\eta_3 -\epsilon_2\eta_2=0$. 

In conclusion, all holomorphic functions $E_{\CY_-}, J_{\CY_-}$ have been completely determined  up to 
certain ambiguous phases which can be set to $\pm 1$ by field redefinitions. The final 
results are summarized in the following table 
\be\label{eq:EJfinal} 
\bal 
 & \CY_-\qquad & & E_{\CY_-} & & J_{\CY_-}&  \\
 & \CX_{-} & & -\Phi_+\Gamma_+& & 
 [\CA_{1+},\CA_{2+}]+\CI_+\CJ_+ &  \\
 &\Omega_-& & \Phi_+\CB_{2+}-\CA_{2+}\Phi_+ & & - \Gamma_+\CA_{1+}& \\
 & \Psi_-& & \CB_{2+}\Gamma_+ -\Gamma_+\CA_{2+}& & \CA_{1+}\Phi_+& \\
 & \Lambda_-& & \CJ_+\Phi_+& &\Gamma_+\CI& \\
\eal 
\ee
Then the  total  potential energy of the  quantum mechanical effective action is  
\be\label{eq:effactB} 
U = U_{gauge} + U_{D} + U_{E} + U_J
\ee
where $U_{gauge}$ is the potential energy determined by gauge couplings, 
\be\label{eq:gauge} 
\bal 
 U_{gauge} =\ & |[\sigma_1,A_1]|^2 + |[\sigma_1,A_2]|^2+ |[\sigma_2,B_2]|^2+|\sigma_1I|^2 +|J\sigma_1|^2 \\
& + |\sigma_1 f - f \sigma_2|^2 + |\sigma_2g -g \sigma_1|^2, \\
\eal 
\ee
$U_D$ is the D-term contribution 
 \be\label{eq:Dterm} 
\bal 
U_D=\ & \left([A_1,A_1^\dagger] + [A_2,A_2^\dagger] +  II^\dagger - J^\dagger J 
+ff^\dagger - g^\dagger g -\zeta_1\right)^2 \\ & + 
 \left([B_2,B_2^\dagger] + g g^\dagger - f^\dagger f - \zeta_2 \right)^2,\\
 \eal 
 \ee
 and 
\be\label{eq:effactA} 
\bal 
 U_E+U_J=\ &  
 |[A_1,A_2]+IJ|^2 + |fg|^2 + |A_1f|^2 + |gA_1|^2 \\ & 
 +  |A_2f-fB_2|^2 + |gA_2-B_2g|^2 
+ |Jf|^2 + |gI|^2\\
\eal 
\ee
are the $E$ and $J$ term contributions. 

The supersymmetric ground states of the resulting quantum-mechanical 
system are obtained in the Born-Oppenheimer approximation by 
quantization of the moduli space of classical supersymmetric flat directions. 
As usual in supersymmetric theories, this approximation yields an exact 
count of such states. 
The geometry of the resulting moduli space will be studied in the next section.

\section{Moduli space of flat directions and enhanced ADHM data}\label{sectionthree}

The main goal of this section is to analyze the geometry of the moduli space
of
supersymmetric flat directions of the quantum mechanical potential
\eqref{eq:effactA}. It will be shown below that, for generic values of the
FI parameters, such moduli space is isomorphic to the moduli space of stable
representations of a quiver with relations, called the enhanced ADHM quiver.
It will be also shown that, in a certain stability chamber, this
moduli space admits a geometric interpretation in terms of
framed torsion free sheaves on the projective plane.

Summarizing the results of the previous section, the D2-brane effective action has
been constructed by dimensional
reduction of a $(0,2)$ model with field content given by the quiver diagram
\eqref{eq:zerotwoquiverC} and interactions
given by \eqref{eq:EJfinal}. The space of constant field configurations
$(A_1,A_2,I,J,B_2,f,g,\sigma_1,\sigma_2)$ is the vector space
\be\label{eq:constfield}
\bal
& \mathrm{End}(V_1)^{\oplus 2} \oplus \mathrm{Hom}(W,V_1)\oplus
\mathrm{Hom}(V_1,W)\, \oplus \\
& \mathrm{End}(V_2) \oplus \mathrm{Hom}(V_1,V_2) \oplus
\mathrm{Hom}(V_2,V_1) \oplus {\mathfrak u}(V_1) \oplus
{\mathfrak u}(V_2)\\
\eal
\ee
where $V_1$, $V_2$ and $W$ are complex vector spaces equipped with hermitian
inner products. The moduli space of flat directions is the moduli space of
gauge equivalence
classes of solutions to the zero-energy equations
 \be\label{eq:flatdirA}
 \bal
& [A_1,A_1^\dagger] + [A_2,A_2^\dagger] +  II^\dagger - J^\dagger J
+ff^\dagger - g^\dagger g =\zeta_1,\\
& [B_2,B_2^\dagger] + g g^\dagger - f^\dagger f = \zeta_2,\\
\eal
\ee
 \be\label{eq:flatdirB}
\bal
& [A_1,A_2]+IJ= 0,\qquad Jf=0,\qquad gI=0,\qquad A_1f=0, \qquad gA_1=0\\
& A_2f-fB_2=0,\qquad gA_2-B_2g=0,\qquad fg=0,\\
 \eal
 \ee
 \be\label{eq:flatdirC}
\bal
& [\sigma_1,A_i]=0,\qquad  [\sigma_2,B_2]=0,\qquad
\sigma_1I=0,\qquad J\sigma_1=0, \\
&  \sigma_1f-f\sigma_2=0,\qquad
g\sigma_1-\sigma_2g=0\\
\eal
\ee
derived from \eqref{eq:effactB}.
Two solutions are gauge equivalent if they are related by the natural
action of the gauge group $U(V_1)\times U(V_2)$
on the space \eqref{eq:constfield}. The resulting moduli space can be
naturally
identified with a moduli space of quiver representations, as presented
below.

\subsection{Enhanced ADHM Quiver}\label{sectionthreeone}
The enhanced ADHM quiver is the quiver with relations defined by the
following diagram
\be\label{eq:enhquiverA}
\xymatrix{
 \ar@(ul,dl)|{\beta} e_2 \ar@/^/[rr]|{\phi} & & e_1 \ar@/^/[ll]|{\gamma}
  \ar@(ur,ul)|{\alpha_1}\ar@(dl,dr)|{\alpha_2} \ar@/^/[rr]|{\eta}
&&  e_\infty ,\ar@/^/[ll]|{\xi}\\}
\ee
and ideal of relations being generated by
\be\label{eq:relationsA}
\bal
\alpha_1\alpha_2-\alpha_2\alpha_1+\xi\eta, &\qquad
\alpha_1\phi, \qquad \alpha_2\phi - \phi \beta, \qquad
\eta \phi,  \qquad \gamma\xi\\
& \phi\gamma,\qquad \gamma\alpha_1,\qquad \gamma\alpha_2-\beta\gamma. \\
\eal
\ee
Note that omitting the vertex $e_2$ and all above relations except the first
one,
one obtains the usual ADHM quiver.

A representation $\CR$ of the enhanced ADHM quiver in the category of
complex
vector spaces is given by a triple $(V_1,V_2,W)$ of vector spaces
assigned to the vertices $(e_1,e_2,e_\infty)$ and linear maps
$(A_1,A_2,I,J,B,f,g)$
assigned to the arrows $(\alpha_1,\alpha_2,\xi,\eta,\beta,\phi,\gamma)$
respectively, and
satisfying the relations \eqref{eq:relationsA}. The numerical type of a
representation
is the triple $(\mathrm{dim}(W),\mathrm{dim}(V_1), \mathrm{dim}(V_2))
\in (\IZ_{\geq 0})^3$. A morphism between two such representations $\CR$ and
$\CR'$ is
a triple $(\xi_1,\xi_2,\xi_\infty)$ of linear maps between the vector spaces
assigned to the nodes $(e_1,e_2,e_\infty)$, respectively, satisfying obvious
compatibility conditions
with the morphisms attached to the arrows. This defines an abelian category
of
quiver representations. Note that this abelian category contains the abelian
category of representations of the ADHM quiver as the full subcategory of
representations
with $n_2=0$.

A framed representation of the enhanced ADHM quiver with type $(r,n_1,n_2)\in (\IZ_{\geq 0})^3$
is a pair $(\CR,h)$ consisting of a representation $\CR$ and an isomorphism
$h:W{\buildrel \sim \over \longto } \IC^r$.
Two framed representations $(\CR,h)$ and $(\CR',h')$ are isomorphic if there
is an
isomorphism of the form
$(\xi_1,\xi_2,\xi_\infty):\CR{\buildrel \sim \over \longto}\CR'$ such that
$h'\xi_\infty=h$.

In order to construct moduli spaces of framed representations of the
enhanced ADHM quiver,
one has to introduce suitable stability conditions. By analogy with
\cite{King},
a stability condition will be defined by a triple
$\theta=(\theta_1,\theta_2,\theta_\infty)\in \IQ^3$
satisfying the relation
\be\label{eq:thetarelation}
n_1\theta_1+n_2\theta_2+r\theta_\infty=0.
\ee
A representation $\CR$ of numerical type $(r,n_1,n_2)\in (\IZ_{>0})^3$
will be called $\theta$-(semi)stable if the following conditions hold
\begin{itemize}
\item[$(i)$]
Any subrepresentation $\CR'\subset \CR$ of numerical type $(0,n_1',n_2')$
 satisfies
\be\label{eq:thetastabA}
n_1'\theta_1+n_2'\theta_2 \ (\leq)\ 0.
\ee
\item[$(ii)$]
Any subrepresentation $\CR'\subset \CR$ of numerical type $(r,n_1',n_2')$
   satisfies
\be\label{eq:thetastabA}
n_1'\theta_1+n_2'\theta_2 + r\theta_\infty \ (\leq)\ 0.
\ee
\end{itemize}
We emphasize that the above definition does not coincide with the one
considered by King in
\cite[Section 3]{King} because only subrepresentations with $r'=0,r$ 
are considered in the stability condition. However, as we shall see in the next subsection, it
plays essentially the same role.

Note also that $\theta$-stability has the Harder-Narasimhan, respectively
Jordan-H\"older
property since the abelian category of quiver representations is noetherian
and artinian.
Two $\theta$-semistable representation with identical dimension vectors will
be called
$S$-equivalent if their associated graded representations with respect to
the
Jordan-H\"older filtration are isomorphic.

Let $(r,n_1,n_2)\in (\IZ_{>0})^3$ be a fixed dimension vector.
Note that the space of stability parameters
$\theta =(\theta_1,\theta_2,\theta_\infty)\in \IQ^3$
satisfying $n_1\theta_1+n_2\theta_2+r\theta_\infty=0$
can be naturally identified with the $(\theta_1,\theta_2)$-plane $\IQ^2$,
after solving for $\theta_\infty$.
Such a parameter $\theta$
will be  called critical of type $(r,n_1,n_2)$ if the set of strictly
$\theta$-semistable representations
$\CR$ with dimension vector $(r,n_1,n_2)$ is non-empty. If this set is
empty,
$\theta$ will be called generic. Then it is easy to prove that, for a fixed
dimension vector $(r,n_1,n_2)\in (\IZ_{>0})^3$, the set of critical
stability parameters consists of of finitely
many lines in the $(\theta_1,\theta_2)$-plane.

The following lemma establishes the existence of generic stability
parameters for any given dimension vector $(r,n_1,n_2)$.

% special stability chambers in the $(\theta_1,\theta_2)$-plane

\begin{lemm}\label{thetachamber}
Suppose $\theta_2>0$ and $\theta_1+n_2\theta_2 <0$ for
some fixed $(r,n_1,n_2)\in (\IZ_{> 0})^3$. Then a
representation $\CR$ is $\theta$-semistable if and only if it is
$\theta$-stable
and if and only if the following conditions are satisfied
\begin{itemize}
\item[(S.1)] $f: V_2\to V_1$ is injective and $g: V_1\to V_2$ is identically
zero.
\item[(S.2)] The data
$\CA=(V_1,W,A_1,A_2,I,J)$ satisfies the ADHM stability condition,
that is
 there is no proper nontrivial subspace $0\subset V_1'\subset V_1$
preserved by $A_1,A_2$ and containing the image of $I$.
\end{itemize}
\end{lemm}

{\it Proof.} Under the assumptions of lemma (\ref{thetachamber}) let
$\CR$ be a $\theta$-semistable representation. Suppose $f$ is not injective.
Then it is straightforward to check that $\mathrm{\ker}(f)\subset V_2$
is preserved by $B_2$, therefore it determines a subrepresentation of $\CR$
with $n_1'=0$, $r'=0$. The semistability condition yields
\[
\theta_2 \, \mathrm{dim}(\mathrm{Ker}(f)) \leq 0
\]
which leads to a contradiction if $\mathrm{dim}(\mathrm{Ker}(f)) >0$.
Therefore $f$ must be injective, and relation $fg=0$ implies $g=0$.

Similarly, if condition $(S.2)$ is not satisfied by some proper nontrivial
subspace
$0\subset V_1'\subset V_1$, the data
\[
\CR'=(V_1',0,W,A_1|_{V_1'}, A_2|_{V_1'}, I, J|_{V_1'}, 0,0)
\]
determines a proper nontrivial subrepresentation of $\CR$ with $r'=r$ so
that
\[
n_1'\theta_1+n_2'\theta_2 + r\theta_\infty = (n_1'-n_1)\theta_1>0.
\]
This is again a contradiction.

Next let $\CR$ be a representation satisfying conditions $(S.1)$, $(S.2)$,
and suppose $\CR'\subset \CR$ is a nontrivial proper subrepresentation of
$\CR$.
Note that $g'=0$ since $g=0$. There are two cases,  $r'=r$ and $r'=0$.

Suppose $r'=r$.  Then $(S.2)$ implies that $I$ is not identically zero,
hence $n_1'>0$.
If $n_1'<n_1$, the data $\CA'=(V_1',A_1',A_2',I',J')$ would violate
condition
$(S.2)$. Therefore $n_1=n_1'$. Since $\CR'$ has to be a proper
subrepresentation,
$n_2'<n_2$. Then
\[
n_1'\theta_1+n_2'\theta_2 +r\theta_\infty =
 (n_2'-n_2)\theta_2<0.
\]

Now suppose $r'=0$.
Note that $n_1'=0$ implies that $V_2'\subset \mathrm{Ker}(f)=0$, hence
$n_2'=0$ as well. This is impossible since $\CR'$ is assumed nontrivial.
Therefore $n_1'\geq 1$, and
\[
n_1'\theta_1+n_2'\theta_2 \leq \theta_1 + n_2\theta_2 <0
\]
using the conditions of lemma (\ref{thetachamber}).

\hfill $\Box$

In the following, a representation $\CR$ of the enhanced ADHM quiver
will be called stable if it satisfies conditions $(S.1)$, $(S.2)$ of lemma
(\ref{thetachamber}).

\subsection{Moduli spaces}\label{sectionthreetwo}

Moduli spaces of $\theta$-semistable framed quiver representations
will be constructed employing GIT techniques, by analogy to \cite{King}.
Since framed quiver moduli of the type considered here
do not seem to be treated previously in the literature, the details will be
presented below
for completeness.

Let $V_1,V_2,W$ be vector spaces of dimensions $n_1,n_2,r \in \IZ_{>0}$
respectively.
Let \[
\bal
{\mathbb X}(r,n_1,n_2)= & \mathrm{End}(V_1)^{\oplus 2} \oplus \mathrm{Hom}(W,V_1)\oplus
\mathrm{Hom}(V_1,W)\oplus \\
& \mathrm{End}(V_2) \oplus \mathrm{Hom}(V_1,V_2) \oplus
\mathrm{Hom}(V_2,V_1). \\
\eal
\]
and note that there is a natural $G=GL(V_1)\times GL(V_2)$ action on
${\mathbb X}(r,n_1,n_2)$
given by
\[
\bal
(g_1,g_2) \times (A_1,A_2, & I,J,B_2,f,g) \longto \\
& (g_1A_1g_1^{-1},g_1A_2g_1^{-1}, Jg_1^{-1}, g_1I,
g_2B_2g_2^{-1},g_1fg_2^{-1},g_2gg_1^{-1}).\\
\eal
\]
The closed points of ${\mathbb X}(r,n_1,n_2)$ will be denoted by
${\sf x}=(A_1,A_2,I,J,B_2,f,g)$, and the action of $(g_1,g_2)\in G$ on a
point ${\sf x}\in {\mathbb X}$ will be denoted by $(g_1,g_2)\cdot {\sf x}$.
The stabilizer of a given point ${\sf x}$ will be denoted by $G_{\sf
x}\subset G$.
Moreover, let ${\mathbb X}_0(r,n_1,n_2)\subset {\mathbb X}$ denote the subscheme
defined by the
algebraic equations \eqref{eq:flatdirB}. Obviously, ${\mathbb X}_0(r,n_1,n_2)$ is
preserved by the $G$-action.

Note also each representation $\CR=(V_1,V_2,W,A_1,A_2,I,J,B_2,f,g)$
corresponds to a unique point ${\sf x}=(A_1,A_2,I,J,B_2,f,g)$
in ${\mathbb X}_0$; two framed representations are isomorphic if and only if
the corresponding points in ${\mathbb X}_0(r,n_1,n_2)$ are in the same $G$-orbit.

Next, recall some standard facts on GIT quotients for a reductive algebraic
group $G$ acting on a vector space ${\mathbb X}(r,n_1,n_2)$ \cite[Section 2]{King}.
Given an algebraic  character
$\chi:G\to \IC^\times$ one has the following notion of
$\chi$-(semi)stability.
\begin{itemize}
\item[$(a)$] A point ${\sf x}_0$ is called $\chi$-semistable if there exists
a polynomial function $p({\sf x})$ on ${\mathbb X}(r,n_1,n_2)$ satisfying
$p((g_1,g_2)\cdot
{\sf x}) = \chi(g_1,g_2)^l p({\sf x})$ for some $l\in \IZ_{\geq 1}$, so that
$p({\sf x}_0)\neq 0$.
\item[$(b)$] A point ${\sf x}_0$ is called $\chi$-stable if there
exists a polynomial function $p({\sf x})$ as in $(a)$ above so that
 $\mathrm{dim}(G\cdot {\sf x}_0) = \mathrm{dim}(G/\Delta)$,
 where ${\Delta \subset G}$ is the subgroup acting trivially on ${\mathbb
X}(r,n_1,n_2)$.
 and the action of $G$ on $\{{\sf x} \in {\mathbb X}(r,n_1,n_2)\, |\, p({\sf x})\neq
0\}$ is closed.
 \end{itemize}
This definition can be reformulated as follows.
Let $G$ act on the direct product ${\mathbb X}_0(r,n_1,n_2)\times
\IC$ by
\[
(g_1,g_2)\times ({\sf x}, z) \to ((g_1,g_2)\cdot {\sf x},
\chi(g_1,g_2)^{-1}z)
\]
Then according to \cite[Lemma 2.2]{King}, ${\sf x} \in {\mathbb X}(r,n_1,n_2)$ is
$\chi$-semistable if and only if
the closure of the orbit $G\cdot({\sf x},z)$ is disjoint
from the zero section ${\mathbb X}(r,n_1,n_2)\times \{0\}$, for any $z\neq 0$.
Moreover ${\sf x}$ is $\chi$-stable if and only if the orbit $G\cdot({\sf
x},z)$ is closed
in complement of the zero section, and the stabilizer $G_{(\sf x,z)}$ is a
finite index subgroup of $\Delta$.

One can form the quasi-projective scheme:
$$ \CN_\theta^{ss}(r,n_1,n_2) = {\mathbb X}_0(r,n_1,n_2) /\!/_{\chi} G :=
{\rm Proj}\left( \oplus_{n\geq 0} A({\mathbb X}_0(r,n_1,n_2))^{G,\chi^n}
\right)~~, $$
where
$$ A({\mathbb X}_0(r,n_1,n_2))^{G,\chi^n} :=
\left\{ f\in A({\mathbb X}_0(r,n_1,n_2)) ~|~ f(g\cdot x)= \chi(g)^n f(x)
~\forall g\in G \right \} ~~.$$
Clearly, $\CN_\theta^{ss}(r,n_1,n_2)$ is projective over ${\rm
Spec}\left({\mathbb X}_0(r,n_1,n_2))^{G}\right)$,
and it is quasi-projective over $\IC$. Geometric Invariant Theory tells us
that $\CN_\theta^{ss}(r,n_1,n_2)$ is
the space of $\chi$-semistable orbits; moreover, it contains an open
subscheme $\CN_\theta^{s}(r,n_1,n_2)\subseteq \CN_\theta^{ss}(r,n_1,n_2)$
consisting of $\chi$-stable orbits.

Then the following holds by analogy with \cite[Prop. 3.1, Thm. 4.1]{King}.
Again
the details of the proof are given below for completeness.

\begin{prop}\label{thetamoduli}
Suppose $\theta=(\theta_1,\theta_2)\in \IZ^2$, and let
$\chi_\theta:G\to \IC^\times$ be the character
\[
\chi_\theta(g_1,g_2)=
\mathrm{det}(g_1)^{-\theta_1}\mathrm{det}(g_2)^{-\theta_2}.
\]
Then a representation $\CR=(V_1,V_2,W,A_1,A_2,I,J,B_2,f,g)$ of
an enhanced ADHM quiver, of dimension vector $(r,n_1,n_2)\in (\IZ_{>0})^3$,
is $\theta$-(semi)stable if and only if the corresponding closed point ${\sf
x} \in
{\mathbb X}_0$ is $\chi_\theta$-(semi)stable.
\end{prop}

It follows that $\CN_\theta^{ss}(r,n_1,n_2)$ parameterizes S-equivalence
classes of $\theta$-semistable framed representations, while
$\CN_\theta^{s}(r,n_1,n_2)$
parameterizes isomorphism classes of $\theta$-stable framed representations.

{\it Proof.}
First, we prove that if ${\sf x}\in {\mathbb X}$ is
$\chi_\theta$-semistable, then the corresponding representation $\CR$ is
$\theta$-semistable.
Suppose that there exists a nontrivial proper subrepresentation $0\subset
\CR'\subset \CR$ with either $r'=0$ or $r'=r$  so that
\[
n_1'\theta_1+n_2'\theta_2 + r'\theta_\infty >0.
\]

Let us first consider the case $r'=0$. Since
$\CR'=(V_1',V_2',\{0\},A_1',A_2',I',J',B_2',f',g')$ is a subrepresentation
of $\CR$, then $V_1'$ and $V_2'$ can be regarded as subspaces of $V_1$ and
$V_2$, respectively, and it follows that
\be\label{eq:subrepA}
\bal
f(V_2') \subseteq V_2',& \qquad g(V_2')\subseteq V_1', \qquad
A_i(V_1')\subseteq V_1', \\ & B_2(V_2') \subseteq V_2', \qquad J(V_1')=0,\\
\eal
\ee
for $i=1,2$.
Then there exist direct sum decompositions $V_1\simeq V_1'\oplus V_1''$,
$V_2\simeq V_2'\oplus V_2''$ such that the linear maps $A_1$, $A_2$, $B_2$,
$f$, and $g$ have
block decomposition of the form
\be\label{eq:blockformsA}
\bal
 \left[\begin{array}{cc} \ast & \ast \\ 0 & \ast  \\
 \end{array}\right]
 \eal
\ee
while $I$, $J$ have block decompositions of the form
\be\label{eq:blockformsB}
\bal
 I=\left[\begin{array}{c} \ast  \\  \ast  \\
 \end{array}\right], \qquad
J= \left[\begin{array}{cc} 0 &  \ast  \\
 \end{array}\right].
 \eal
\ee
Consider a one-parameter subgroup of $G$ of the form
\[
g_1(t)=  \left[\begin{array}{ll} t1_{V_1'} & 0 \\ 0 &  1_{V_1''}\\
 \end{array}\right] ,\qquad
 g_2(t) = \left[\begin{array}{ll} t1_{V_2'} & 0 \\ 0 &  1_{V_2''}\\
 \end{array}\right] .
 \]
It follows that the linear maps
$(A_1(t),A_2(t),I(t),J(t),B_2(t),f(t),g(t))=(g_1(t),g_2(t))\cdot {\sf x}$
have block decompositions of the form
\be\label{eq:blockformsC}
\bal
 \left[\begin{array}{cc} \ast & t \ast \\ 0 & \ast  \\
 \end{array}\right] ,
 \eal
\ee
and
\be\label{eq:blockformsD}
\bal
 I^t=\left[\begin{array}{c} t\ast  \\  \ast  \\
 \end{array}\right], \qquad
J^t= \left[\begin{array}{cc} 0 &  \ast  \\
 \end{array}\right].
 \eal
\ee
At the same time, $\chi_\theta(g_1(t),g_2(t))^{-1}z=
t^{n_1'\theta_1+n_2'\theta_2}z$,
with $n_1'\theta_1+n_2'\theta_2>0$.
Therefore the limit of $(g_1(t),g_2(t))\cdot ({\sf x}, z)$ as $t\to 0$ is a
point
on the zero section, which contradicts $\chi_\theta$-semistability.

Suppose ${\sf x}$ is $\chi_\theta$-stable but $\CR$ is not $\theta$-stable.
Then the previous argument
shows that $\CR$ must be $\theta$-semistable, therefore there must exist a
nontrivial
proper subrepresentation $0\subset \CR'\subset \CR$, $r'=0$ or $r'=r$, so
that
\[
n_1'\theta_1+n_2'\theta_2 + r'\theta_\infty =0.
\]
Since the orbit $G\cdot({\sf x}, z)$ must be closed in the complement of
the zero section for any $z\neq 0$ it follows  that the block decompositions
\eqref{eq:blockformsA} must be diagonal, and the upper block in the
decomposition
of $I$ in \eqref{eq:blockformsB} must be trivial. Otherwise the limit of
$(g_1(t),g_2(t))
\cdot ({\sf x},z)$ exists, but does not belong to the $G$-orbit through
$({\sf x}, z)$.  However, this implies that the
one-parameter subgroup $(g_1(t),g_2(t))$ stabilizes $({\sf x}, z)$. Since
the kernel
$\Delta$ of the representation of $G$ on ${\mathbb X}$ is trivial, this
contradicts
the $\chi_\theta$-stability assumption. Therefore $\CR$ must be
$\theta$-stable.

Next, consider the case $r'=r$. As in the previous case, it follows that
\be\label{eq:subrepA} \bal
f(V_2') \subseteq V_2',& \qquad g(V_2')\subseteq V_1', \qquad
A_i(V_1')\subseteq V_1', \\ & B_2(V_2') \subseteq V_2', \qquad I(W)\subseteq
V_1',\\
\eal
\ee
for $i=1,2$. Therefore there exist  direct sum decompositions $V_1\simeq
V_1'\oplus V_1''$,
$V_2\simeq V_2'\oplus V_2''$ such that the linear maps $(A_1,A_2,B_2,f,g)$
have
block decomposition of the form \eqref{eq:blockformsA} while $I,J$ have
block form
decompositions of the form
\be\label{eq:blockformsE}
\bal
 I=\left[\begin{array}{c} \ast  \\  0 \\
 \end{array}\right], \qquad
J= \left[\begin{array}{cc} \ast &  \ast  \\
 \end{array}\right].
 \eal
\ee
Consider a one-parameter subgroup of $G$ of the form
\[
g_1(t)=  \left[\begin{array}{ll} 1_{V_1'} & 0 \\ 0 &  t^{-1}1_{V_1''}\\
 \end{array}\right] ,\qquad
 g_2(t) = \left[\begin{array}{ll} t1_{V_2'} & 0 \\ 0 &  t^{-1}1_{V_2''}\\
 \end{array}.\right]
 \]
 Then the linear maps $(A_1^t,A_2^t,B_2^t,f^t,g^t)$ in $(g_1(t),g_2(t))\cdot
{\sf x}$
have block decompositions of the form \eqref{eq:blockformsC} and
$(I^t,J^t)$ have block decompositions
\be\label{eq:blockformsF}
\bal
 I^t=\left[\begin{array}{c} \ast  \\  0  \\
 \end{array}\right], \qquad
J^t= \left[\begin{array}{cc} \ast &  t\ast  \\
 \end{array}\right].
 \eal
\ee
Since $\chi_\theta(g_1(t),g_2(t))^{-1} z =
t^{(n_1'-n_1)\theta_1+(n_2'-n_2)\theta_2}z$,
this leads again to a contradiction.

Suppose ${\sf x}$ is $\chi_\theta$-stable, but $\CR$ is not $\theta$-stable.
Then, as above,
it follows that the block decompositions \eqref{eq:blockformsA} must be
diagonal, and
the left block in the decomposition of $J$ in \eqref{eq:blockformsD} must be
trivial.
This again implies that ${\sf x}$ has nontrivial stabilizer, leading to a
contradiction.

The proof of the converse statement is very similar, the details being left
to the reader.

\hfill $\Box$

As observed above Lemma \ref{thetachamber}, for fixed dimension vector
$(r,n_1,n_2)\in (\IZ_{>0})^3$, the space of stability parameters $\theta$
can be naturally identified with the $(\theta_1,\theta_2)$-plane and there
is a
critical set of lines through the origin dividing it into  finitely many
stability chambers.
All moduli spaces associated to stability parameters within a chamber are
canonically isomorphic and do not contain strictly semi-stable points.

Lemma \ref{thetachamber} shows that there is a special stability chamber,
determined by the inequalities $\theta_2>0$, $\theta_1+n_2\theta_2<0$,
within which $\theta$-semistability is equivalent to $\theta$-stability and
to conditions $(S.1)$, $(S.2)$ stated in Lemma \ref{thetachamber}.
Framed representations of the enhanced ADHM quiver satisfying conditions
$(S.1)$, $(S.2)$ will simply be called stable, and their
moduli space will be denoted by $\CN(r,n_1,n_2)$.

\begin{theo}\label{flatdirthm}
Let $(r,n_1,n_2)\in (\IZ_{>0})^3$ be a fixed dimension vector and
$\theta=(\theta_1,\theta_2,\theta_\infty)\in \IZ^2 \times \IQ$
be a generic stability parameter.
Then the set of gauge equivalence classes of solutions to equations
\eqref{eq:flatdirA}-\eqref{eq:flatdirC} with $\zeta_1=\theta_1$ and
$\zeta_2=-\theta_2$
is a complex quasi-projective
scheme isomorphic to $\CN_\theta^s(r,n_1,n_2)$.
\end{theo}

{\it Proof.}
The two equations in \eqref{eq:flatdirA} are obviously moment map equations
for the
natural hamiltonian $U(V_1)\times U(V_2)$-action on the vector space
\[
\bal
{\mathbb X}(r,n_1,n_2)= & \mathrm{End}(V_1)^{\oplus 2} \oplus \mathrm{Hom}(W,V_1)\oplus
\mathrm{Hom}(V_1,W)\oplus \\
& \mathrm{End}(V_2) \oplus \mathrm{Hom}(V_1,V_2) \oplus
\mathrm{Hom}(V_2,V_1). \\
\eal
\]
The parameters $(\zeta_1,\zeta_2)$ determine the level of the moment map
$\mu:{\mathbb X}(r,n_1,n_2) \to {\mathfrak u}(V_1)^*\oplus {\mathfrak u}(V_2)^*$.
Standard results imply that for generic $(\theta_1,\theta_2)\in \IZ^2$, the
symplectic K\"ahler quotient $\mu^{-1}(-\theta_1,-\theta_2)/ U(V_1)\times
U(V_2)$,
is isomorphic to the GIT quotient ${\mathbb X}_0(r,n_1,n_2) /\!/_{\chi} G$,
where $\chi:G\to \IC^\times$ is a character of the form
\[
\chi(g_1,g_2)= \mathrm{det}(g_1)^{-\theta_1}\mathrm{det}(g_2)^{-\theta_2}.
\]

As it was observed below Proposition \ref{thetamoduli}, the GIT quotient
${\mathbb X}_0(r,n_1,n_2) /\!/_{\chi} G$ is isomorphic to the moduli space
of
S-equivalence classes of $\theta$-semistable quiver representations
$\CN^{ss}_\theta(r,n_1,n_2)$. For generic $\theta$ there are no strictly
semistable
representations by Lemma \ref{thetachamber}, hence
$\CN^{ss}_\theta(r,n_1,n_2)=\CN^{s}(r,n_1,n_2)$.
In conclusion, the symplectic quotient
$\mu^{-1}(-\theta_1,-\theta_2)/U(V_1)\times U(V_2)$ is
isomorphic to the moduli space $\CN^{s}_\theta(r,n_1,n_2)$.

Finally, note that equations \eqref{eq:flatdirC} imply that the triple
$(\mathrm{exp}(\sigma_1),\mathrm{exp}(\sigma_2),1_W)$ is an endomorphism of
the enhanced ADHM quiver representation \linebreak
$\CR=(V_1,V_2,W,A_1,A_2,I,J,B_2,f)$ preserving the framing $h:W{\buildrel
\sim \over \longto} \IC^r$. However, the proof of Proposition
(\ref{thetamoduli}) implies that a nontrivial  endomorphism of a stable
framed representation must be the identity.
In conclusion, $\sigma_1,\sigma_2$ must be identically $0$ for generic
$\theta$.

\hfill $\Box$

In particular, it follows from the proof above and from Lemma
\ref{thetachamber} that if $\zeta_2<0$ and $\zeta_1+n_2\zeta_2>0$, then the
moduli space of flat directions is isomorphic to $\CN(r,n_1,n_2)$.

For further reference, note that if $\CR=(V_1,V_2,W,A_1,A_2,I,J,B_2,f)$ is a
stable framed representation of type $(r,n_1,n_2)\in \IZ_{>0}^3$ with
$n_1>n_2$, the linear maps
$(A_1,A_2,I,J)$ yield linear maps
\[
{\widetilde A}_i:V_1/\mathrm{Im}(f) \to V_1/\mathrm{Im}(f), \qquad
{\widetilde I}: W\to V_1/\mathrm{Im}(f),\qquad
{\widetilde J}: V_1/\mathrm{Im}(f)\to W
\]
with $i=1,2$, which satisfy the ADHM relation
\[
[{\widetilde A}_1, {\widetilde A}_2]+ {\widetilde I}{\widetilde J} =0.
\]
Moreover, it is not difficult to check that the resulting ADHM data
$(V,W,{\widetilde A}_1, {\widetilde A}_2,{\widetilde I},{\widetilde J})$,
where
$V=V_1/\mathrm{Im}(f)$ satisfies the ADHM stability condition $(S.2)$.

\begin{lemm}\label{projlemma}
Suppose $n=n_1-n_2>0$ and let $V_2$ be a complex vector space of dimension
$n_2\in \IZ_{>0}$. Let also $\CM(r,n)$ denote the moduli space of stable ADHM data of
type $(n,r)\in (\IZ_{>0})^2$.
Then there is a surjective morphism ${\mathfrak q} : \CN(r,n_1,n_2) \to
\CM(r,n)$ mapping a the isomorphism class of the stable framed representation
$\CR=(V_1,V_2,W,A_1,A_2,I,J,B_2,f)$ to isomorphism class of the ADHM data
$(V,W,{\widetilde A}_1, {\widetilde A}_2,{\widetilde I},{\widetilde J})$
constructed above.
\end{lemm}

{\it Proof.} The existence of the morphism ${\sf q}$ of moduli spaces
follows from
repeating the above construction for flat families of quiver
representations.

In order to prove its surjectivity, start with a stable ADHM data
$(V,W,{\widetilde A}_1, {\widetilde A}_2,{\widetilde I},{\widetilde J})$ of
type $(n,r)$
and $B_2\in\mathrm{End}(V_2)$, and set
\[
V_1=V_2\oplus V, ~~{\rm and}~~ \qquad f = \left[\begin{array}{c}
1_{V_2}\\ 0 \end{array}\right].
\]

Now let $A_1,A_2\in\mathrm{End}(V_1)$, $I\in\mathrm{Hom}(W,V_1)$, and
$J\in\mathrm{Hom}(V_1,W)$ be of the following form
$$ A_1 = \left[\begin{array}{cc} 0 & A_1' \\ 0 & {\widetilde A}_1
\end{array}\right] ~~,~~
A_2 = \left[\begin{array}{cc} B_2 & A_2' \\ 0 & {\widetilde A}_2
\end{array}\right] $$
$$ I = \left[\begin{array}{c} I' \\ {\widetilde I} \end{array}\right] ~~,~~
J = \left[\begin{array}{cc} 0 & {\widetilde J} \end{array}\right], $$
according to the decomposition $V_1=V_2\oplus V$. To be precise, one has
$A_1',A_2'\in\mathrm{Hom}(V,V_2)$ and $I'\in\mathrm{Hom}(W,V_2)$.

One immediately sees that $A_1f=A_2f-fB_2=Jf=0$, while $[A_1,A_2]+IJ=0$ if
and only if the following auxiliary equation is satisfied:
\begin{equation}\label{eq:auxeq}
A_1'{\widetilde A}_2 - A_2'{\widetilde A}_1 - B_2A_1' + I'{\widetilde J} =
0.
\end{equation}

Clearly, $f:V_2\to V_1$ is injective, and note that $(V_1,W,A_1,A_2,I,J)$
defined above is stable if and only if the following conditions hold:
\begin{itemize}
\item[(i)] at least one of the linear maps $A_1',A_2',I'$ is nontrivial;
\item[(ii)] there is no proper subspace $S\subsetneq V_2$ such that
$A_1'(V), A_2'(V),I'(W)\subset S$ and $B_2(S)\subseteq S$.
\end{itemize}
Indeed, if $A_1'=A_2'=I'=0$, then $V$ is a subspace of $V_1$ which violates
the ADHM stability condition. As for the second condition,
$(V_1,W,A_1,A_2,I,J)$ is not stable if and only if there is a subspace
${\widetilde S}\subsetneq V_1$ which is invariant under $A_1$ and $A_2$, and
contains the image of $I$. Since $(V,W,{\widetilde A}_1, {\widetilde
A}_2,{\widetilde I},{\widetilde J})$ is stable, ${\widetilde S}$ must be of
the form $S\oplus V$, with $S\subsetneq V_2$
nontrivial, $A_1'(V), A_2'(V),I'(W)\subset S$ and $B_2(S)\subseteq S$.

Therefore, in order to prove the surjectivity of the morphism ${\sf q}$ it
is sufficient to
prove that there exist nontrivial solutions of the auxiliary equation
\eqref{eq:auxeq}, so that linear subspaces
$0\subsetneq S \subsetneq V_2$ as in the previous paragraph do not exist.

Choose a basis $\{v_1,\ldots, v_{n_2}\}$ of $V_2$ and let $B_2$ be a
diagonal
matrix with distinct eigenvalues,
$B_2=\mathrm{diag}(\beta_1,\ldots, \beta_{n_2})$,
$\beta_i\neq \beta_j$ for all $i,j=1,\ldots, n_2$, $i\neq j$.
Let $I':W\to V_2$ be a rank one linear map so that its image
is generated by a vector $v=\sum_{i=1}^{n_2} v_i$.
Note that the set $\{v,B(v), \ldots, B^{n_2-1}(v)\}$ is a basis of $V_2$.
Otherwise there would exist a nontrivial linear relation of the form
\[
\sum_{i=1}^{n_2} x_i B^i(v)=0.
\]
Given the above choice for $B_2$, this would imply that the $x_i$ are a
solution of
the linear system
\[
\sum_{i=1}^{n_2} \beta_j^ix_i =0
\]
where $j=1,\ldots, n_2$. However the discriminant of this system
is the Vandermonde determinant $\Delta(\beta_1, \ldots, \beta_{n_2})=
\prod_{i<j}(\beta_j-\beta_i)$, which is nonzero since the $\beta_i$ are
assumed to
be distinct. Therefore all $x_i$ would have to vanish, leading to a
contradiction.
In conclusion, $\{v,B(v), \ldots, B^{n_2-1}(v)\}$ is a basis of $V_2$.
In particular there are no nontrivial proper subspaces $0\subsetneq S
\subsetneq V_2$ preserved by $B_2$ and containing $\mathrm{Im}(I')$.

Having fixed $B_2,I'$ as in the previous paragraph, equation
\eqref{eq:auxeq} is a linear system of $n_2(n_1-n_2)$ linear equations in
the $2n_2(n_1-n_2)$
variables $A_1',A_2'$. Such a system has a $n_2(n_1-n_2)$ dimensional space
of solutions. Any nontrivial solution determines a set $(V_1,W,A_1,A_2,I,J)$
of stable ADHM data.

\hfill $\Box$

  \subsection{Virtual smoothness}\label{sectionthreethree}

The main result of this subsection is the following.

\begin{theo}\label{smoothmoduli}
The moduli space $\CN(r,n_1,n_2)$ of stable framed representations of an
enhanced ADHM quiver
with fixed numerical invariants $(r,n_1,n_2)\in (\IZ_{>0})^3$ is a 
quasi-projective variety equipped with a perfect obstruction theory. 
Moreover, the infinitesimal deformation space and the obstruction space at any closed point  $[\CR]=[(A_1,A_2,I,J,B_2,f)]\in \CN(r,n_1,n_2)$ are isomorphic to the first and second cohomology groups of the complex $\CC(\CR)$ defined below. 
\be\label{eq:defcomplexA}
 \begin{array}{c}
 \mathrm{End}(V_1)\\
 \oplus \\
 \mathrm{End}(V_2)\\
 \end{array}
 {\buildrel d_0\over \longto}
 \begin{array}{c}
 \mathrm{End}(V_1)^{\oplus 2}\\
 \oplus \\
 \mathrm{Hom}(W,V_1) \\
 \oplus\\
 \mathrm{Hom}(V_1,W) \\
 \oplus \\
 \mathrm{End}(V_2) \\
 \oplus \\
 \mathrm{Hom}(V_2,V_1) \\
 \end{array}
 {\buildrel d_1\over \longto}
 \begin{array}{c}
 \mathrm{End}(V_1)^{}\\
 \oplus \\
 \mathrm{Hom}(V_2,V_1)^{\oplus 2} \\
\oplus \\
\mathrm{Hom}(V_2,W)\\
\end{array}
 {\buildrel d_2\over \longto}
 \begin{array}{c}
 \mathrm{Hom}(V_2,V_1) .\\
 \end{array}
  \ee
The four terms have degrees $0,\ldots,3$, and the differentials are
given by
 \[
 d_0(\alpha_1,\alpha_2)^t = ([\alpha_1,A_1], [\alpha_1,A_2], \alpha_1I,
-J\alpha_1,
 [\alpha_2, B_2], \alpha_1f - f\alpha_2)^t
 \]
 \[
 d_1(a_1,a_2,i,j,b_2,\phi)^t =
 ([a_1,A_2]+[A_1,a_2]+Ij+iJ,A_1\phi+a_1f,A_2\phi+a_2f-fb_2-\phi
B_2,jf+J\phi)^t
 \]
 \[
 d_2(c_1,c_2,c_3,c_4)^t = c_1f +A_2c_2-c_2B_2-A_1c_3-Ic_4.
 \]
  \end{theo}

  {\it Proof.} First note that the moduli space of stable framed
representations
  of the  enhanced ADHM quiver \eqref{eq:enhquiverA}
  can be canonically identified with  the moduli space of
  stable framed representations of the following simpler quiver
  \be\label{eq:enhquiverB}
\xymatrix{
 \ar@(ul,dl)|{\beta} e_2 \ar[rr]|{\phi} & & e_1
  \ar@(ur,ul)|{\alpha_1}\ar@(dl,dr)|{\alpha_2} \ar@/^/[rr]|{\eta}
&&  e_\infty \ar@/^/[ll]|{\xi}\\}
\ee
with relations
\be\label{eq:relationsB}
\alpha_1\alpha_2-\alpha_2\alpha_1+\xi\eta, \qquad
\alpha_1\phi, \qquad \alpha_2\phi - \phi \beta, \qquad
\eta \phi.
\ee
For further reference, let $(\rho_1,\ldots, \rho_4)$ denote the generators
\eqref{eq:relationsB} respectively.

The moduli space ${\widetilde \CN}(r,n_1,n_2)$ of stable framed
representations of numerical type $(r,n_1,n_2)\in (\IZ_{>0})^3$
is defined in complete analogy with the moduli space of similar
representations
of the enhanced ADHM quiver \eqref{eq:enhquiverA}.
In particular, a result analogous to Lemma (\ref{thetachamber})
also holds for $\theta$-stable framed representations of
\eqref{eq:enhquiverB}. Namely, if $\theta_1<0$, $\theta_2>0$,
$\theta_1+n_2\theta_2<0$, a framed representation
$(V_1,V_2,W, A_1,A_2, I,J,B_2,f)$ of \eqref{eq:enhquiverB} is
$\theta$-semistable
if and only if it is $\theta$-stable and if and only if $f$ is injective
and the data $(V_1,W,A_1,A_2,I,J)$ satisfies the ADHM stability condition
(S.2).
Finally, there is an obvious morphism ${\widetilde \CN}(r,n_1,n_2)\to
\CN(r,n_1,n_2)$, which
is an isomorphism according to Lemma (\ref{thetachamber}). This isomorphism
will be
used implicitly in the following, making no distinction between stable
framed representations of \eqref{eq:enhquiverA} and \eqref{eq:enhquiverB}.

The truncated cotangent complex of the moduli space ${\widetilde
\CN}(r,n_1,n_2)$ can be determined by a standard computation in deformation
theory. Such an explicit computation has been carried out in a similar
context, see \cite[Sect. 4.1]{modADHM}. To be more precise, the differential
$d_0$ comes from the linearization of the action of $G$ on ${\mathbb X}$,
while the differential $d_1$ is just the linearization of the relations
(\ref{eq:relationsB}). The only new element in the present case is the fact
that the generators
$(\rho_1,\ldots, \rho_4)$ in \eqref{eq:relationsB} satisfy the relation
\[
\rho_1\phi+\alpha_2\rho_2-\rho_2\beta -\alpha_1\rho_3-\xi\rho_4=0.
\]
This ``relation on relations" yields an extra term in the deformation
complex of a framed representation ${\mathcal
R}=(V_1,V_2,W,A_1,A_2,I,J,B_2,f)$ of the quiver \eqref{eq:enhquiverB}, and
the differential $d_2$ is  precisely its linearization.

We conclude that the infinitesimal deformation space of $\CR$ is the first
cohomology group
$H^1(\CC(\CR))$ and the obstruction space is $H^2(\CC(\CR))$. In order to
prove theorem (\ref{smoothmoduli}),
it suffices to show that $ H^i(\CC(\CR))=0$, for $i\in\{0,3\}$,
for any stable framed representation $\CR$. A helpful observation is that
$\CC(\CR)$ can be presented as a cone of a morphism between simpler
complexes as follows.

Let $\CA=(V_1,A_1,A_2,I,J)$ and $\CB=(V_2,B_2)$, and construct the following
complexes of vector spaces:

 $\bullet\ \CC(\CA)$ is the three term complex
  \be\label{eq:defcomplexB}
 \begin{array}{c}
 \mathrm{Hom}(V_1,V_1)\\
  \end{array}
 {\buildrel d_0\over \longto}
 \begin{array}{c}
 \mathrm{End}(V_1,V_1)^{\oplus 2}\\
 \oplus \\
 \mathrm{Hom}(W,V_1) \\
 \oplus\\
 \mathrm{Hom}(V_1,W) \\
  \end{array}
 {\buildrel d_1\over \longto}
 \begin{array}{c}
 \mathrm{End}(V_1,V_1)^{}\\
 \end{array}
   \ee
 where the  terms have degrees $0,1,2$, and the differentials are given by
 \[
 d_0(\alpha_1) = ([\alpha_1,A_1], [\alpha_1,A_2], \alpha_1I, -J\alpha_1)^t
 \]
 \[
 d_1(a_1,a_2,i,j)^t =
 ([a_1,A_2]+[A_1,a_2]+Ij+iJ);
 \]

 $\bullet \ \CC(\CB)$ is the two-term complex
 \be\label{eq:defcomplexC}
 \mathrm{Hom}(V_2,V_2) {\buildrel d_0\over \longto}
 \mathrm{Hom}(V_2,V_2)
 \ee
 with differential
 \[
 d_0(\alpha_2) = [\alpha_2,B_2]
 \]
 and terms in degrees $0,1$;

 $\bullet \ \CC(\CA,\CB)$ is the three term complex
 \be\label{eq:defcomplexD}
  \begin{array}{c}
  \mathrm{Hom}(V_2,V_1) \\
 \end{array}
 {\buildrel d_0\over \longto}
 \begin{array}{c}
  \mathrm{Hom}(V_2,V_1)^{\oplus 2} \\
\oplus \\
\mathrm{Hom}(V_2,W)\\
\end{array}
 {\buildrel d_1\over \longto}
 \begin{array}{c}
 \mathrm{Hom}(V_2,V_1) \\
 \end{array}
  \ee
  where the terms have degrees $0,1, 2$ and the differentials are
  \[
 d_0(\phi) =
 -(A_1\phi,A_2\phi-\phi B_2, J\phi)^t
 \]
 \[
 d_1(c_2,c_3,c_4)^t = -(A_2c_2-c_2B_2-A_1c_3-Ic_4).
 \]
 Abusing notation, the differentials of the above three complexes
 have been denoted by the same symbols $d_0,d_1$. The distinction
 will be clear from the context. Note that $\CC(\CA)$ is the deformation
 complex of the representation $\CA$ of the standard ADHM quiver.

 It is then straightforward to check that the complex
 $\CC(\CR)[1]$ is the cone of the morphism of complexes
 \[
 \bal
 \varrho:\CC(\CA) \oplus \CC(\CB) & \longto \CC(\CA,\CB) \\
  \varrho_0(\alpha_1,\alpha_2)^t & =-(\alpha_1f-f\alpha_2) \\
  \varrho_1(a_1,a_2,i,j,b_2)^t & = -(a_1f,a_2f-fb_2,jf)^t\\
  \varrho_2(c_1) & =-c_1 f.\\
  \eal
 \]
 In particular, there is an exact triangle
 \be\label{eq:extriangle}
 \CC(\CR)\longto\CC(\CA)\oplus \CC(\CB) \longto \CC(\CA,\CB).
 \ee

Next, note that the following vanishing results hold
\be\label{eq:vanishingA}
H^0(\CC(\CA))=0\qquad H^2(\CC(\CA))=0 \qquad H^2(\CC(\CA,\CB))=0.
\ee
if $\CR$ is stable. The first two follow from observing that $\CC(\CA)$ is
just the
deformation complex of a stable ADHM data; the vanishing of $H^0$ and $H^2$
in this case is a
well-known result.

The last vanishing in \eqref{eq:vanishingA} follows from considering the
dual of the differential
$d_1:\CC^1(\CA,\CB)\to \CC^2(\CA,\CB)$. It reads
\[
d_1^\vee : \mathrm{Hom}(V_1,V_2) \to \begin{array}{cc}
\mathrm{Hom}(V_1,V_2)^{\oplus 2}  \\ \oplus \\ \mathrm{Hom}(W,V_2)
\end{array}
\]
\[
d_1^\vee(\psi) = (B_2\psi-\psi A_2,\psi A_1,\psi I)^t .
\]
Suppose $d_1^\vee(\psi)=0$.
Then it is straightforward to check that $\mathrm{Ker}(\psi)$ is preserved
by
$A_1,A_2$ and contains the image of $I$, which implies that
$\mathrm{Ker}(\psi)$
is either $0$ or $V_1$.
If $\psi$ is injective, then $A_1=0$ and $I=0$,  leading to a contradiction.
Therefore $\psi=0$, and $d_1$ is surjective.

Using a similar argument, it is also straightforward to prove that the
morphism
\[
H^0(\CC(\CA))\oplus H^0(\CC(\CB)) {\buildrel H^0(\varrho)\over \longto}
H^0(\CC(\CA,\CB) )
\]
is injective if the stability conditions are satisfied. Then the long exact
cohomology sequence
of the exact triangle \eqref{eq:extriangle} implies that
\be\label{eq:vanishingB}
H^0(\CC(\CR))=0,\qquad H^3(\CC(\CR))=0.
\ee

\hfill$\Box$

 \subsection{Geometric interpretation in terms of 
  framed sheaves}\label{sectionthreefour} 

Let $S$ be a smooth projective surface and $D,D_{\infty}$ smooth irreducible  
divisors 
on $S$ with transverse intersection. According to \cite{BM-framed}, if 
$D_\infty$ is big and nef, 
and $c\in A^\bullet(S)\otimes \IQ$ (the Chow group of $S$),  there is a   
quasi-projective fine moduli scheme 
$\CM(c)$ parametrizing isomorphism classes of pairs $(E,\xi)$, where 
\begin{itemize}
\item $E$ is a torsion free sheaf on $S$ with numerical invariants $\ch(E)=c$;
\item $\xi:E|_{D_\infty} {\buildrel \sim \over \longto} \CO_{D_\infty}^{\oplus 
r}$ 
is an isomorphism of $\CO_{D_\infty}$-modules. 
\end{itemize} 
In particular there exists a universal framed torsion free sheaf $(\CU,
\varepsilon)$ 
on $\CM(c)\times S$, flat over $\CM(c)$. The class $c=(r,c_1,\ch_2)$ will 
satisfy the constraint
$c_1\cdot D_\infty=0$. Under some additional assumptions (e.g., if the 
condition
$(K_S+D_\infty)\cdot D_\infty <0$ holds), the moduli scheme $\CM(c)$ is 
smooth.

We shall consider the functor $\mathbf F_{r,n,d} \colon \mbox{\em Sch}^{op}_{/ \IC} \to \mbox{\em Sets}$
which to any scheme $T$ associates the isomorphism classes of quadruples $(E_T,\xi_T,G_T,g_T)$, where 
\begin{itemize}
\item $E_T$ is a coherent sheaf on $T\times S$, flat over $T$, such that  for all closed points $t\in T$ the sheaf $E_{T,t}=E_{\vert \{t\}\times S}$ is torsion-free and has fixed Chern character $\ch_0=r$, $\ch_1=0$, $\ch_2=-n$;
\item $\xi_T \colon E_{T\times D_\infty} \to \CO^{\oplus r}_{T\times D_\infty}$ is an isomorphism
of $\CO_{D_\infty\times T}$-modules;
\item $G_T$ is a coherent sheaf on $T\times S$, supported on $T\times D$ and flat over $T$, such that  for all closed points $t\in T$, the sheaf $G_{T,t}$ is a skyscraper of fixed length $d\ge 1$, whose support is disjoint from $T\times (D\cap D_{\infty})$;
\item $g_T\colon E_T\twoheadrightarrow G_T$ is a surjective morphism of 
$\CO_{T\times S}$-modules.
\end{itemize}
Two such quadruples $(E_T,\xi_T,G_T,g_T)$ and  $(E'_T,\xi'_T,G'_T,g'_T)$
are considered to be isomorphic if there exist an isomorphism of $\CO_{T\times D}$-modules
$\phi_T\colon E_T \stackrel{\sim}{\to} E'_T$ and an isomorphism of $\CO_{T\times D}$-modules $\psi_T\colon G_T \stackrel{\sim}{\to} G'_T$ such that the diagrams
$$\xymatrix{
E_{T\vert T\times D_\infty } \ar[d]_{\phi_{T\vert T\times D_\infty } } \ar[r]^\sim_{\xi_T} 
& \CO^{\oplus r}_{T\times D_\infty } \\
E'_{T\vert T\times D_\infty }   \ar[ur]^(.4)\sim_{\xi'_T} 
}\qquad
\xymatrix{E_T \ar@{>>}[r]^{g_T} \ar[d]_{\phi_T} & G_T \ar[d]^{\psi_T} \\
E'_T \ar@{>>}[r]^{g'_T} & G'_T}
$$
commute. There is a forgetful natural transformation from $\mathbf F_{r,n,d}$ to the moduli functor represented by $\CM(r,n)$, which simply forgets the data $G_T$ and $g_T$.

The steps leading to the construction of the moduli space $\CM(r,n)$ 
\cite{BM-framed,stpairs,framed} can be easily generalized to get a moduli scheme
$\CM_D(r,n,d)$ which universally represents the functor $\mathbf F_{r,n,d}$. Moreover the above-mentioned forgetful functor induces a projective morphism $\CM_D(r,n,d)\to\CM(r,n)$. However these results can be obtained in a more economical way by noting that  $\mathbf F_{r,n,d}$ is isomorphic to a Quot functor, which is representable by general theory.
For any $d\geq 1$, let $\mathbf Q_{r,n,d}$ be the functor $\mbox{\em Sch}^{op}_{/\CM(r,n)}\to \mbox{\em Sets}$ which associates 
to a scheme $T\to \CM(r,n)$ over $\CM(r,n)$ an isomorphism 
class of  pairs $(F_T, f_T)$ 
where 
\begin{itemize} 
\item $F_T$ is a flat coherent $\CO_{T\times D}$-module with finite support 
over $T$ of relative length $d$, disjoint from $T\times  (D\cap D_\infty)$, and 
\item $f_T : (\CU_D)_T \to F_T$ is a surjective morphism of 
$\CO_{D\times T}$-modules. 
\end{itemize}
Two such quotients $( F_T, g_T)$ $(F'_T, g'_T)$ are isomorphic if there exists an isomorphism $\eta_T : F_T \to F'_T$ such that 
$f'_T = \eta_T\circ  f_T$.  In accordance with Grothendieck's general theory of the Quot scheme, there exists a relative $\CM(r,n)$-scheme 
$\pi:{\CQ}(\CU_D,d)\to \CM(r,n)$  that universally represents the functor $\mathbf Q_{r,n,d}$.

The previously mentioned natural transformation $\mathbf F_{r,n,d}\to 
\mathbf Q_{r,n,d}$ is defined by $(E_T,\xi_T,G_T,g_T)\to (G_T,g_T)$. The inverse transformation is obtained by taking $E_T=\mbox{ker}(g_T)$ and noting that, as consequence of the condition on the support of $G_T$, the framing of the universal sheaf{\hskip4pt}$\CU$ induces a framing $\xi_T$ on $E_T$. As a consequence, we have an isomorphism of $\CM(r,n)$-schemes $\CM_D(r,n,d)\simeq {\CQ}(E_D,d)$.

%
%representing the functor $Sch^{op}_{\CM(r,n)}\to Sets$ 
%The relative {\it Quot}-scheme
%$\CQ(E_D,d)$ is a fine quasi-projective moduli space, that is there exists a 
% universal quotient $\pi_D^*E_D {\longto} 
%G$ where $\pi_D: \CQ(E_D,d)\times D \to \CM(r,n)\times D$ 
%denotes the natural projection.  

Next let $S=\IP^2$ with homogeneous coordinates $[z_0,z_1,z_2]$ 
 and let $D,D_\infty$ be the hyperplanes defined by $z_1=0$ and 
 $z_0=0$, respectively.  Then the moduli space 
$\CM(r,n)$ is isomorphic to the moduli space of stable ADHM data of type $(r,n)$
\cite[Thm. 2.1]{hilblect}. 
Let $\CN(r,n+d,d)$ denote the moduli space of stable representations of 
an enhanced ADHM quiver of type $(n+d,d,r)$. 
 Recall also that Lemma (\ref{projlemma}) proves the existence of a surjective 
 morphism ${\sf q}: \CN(r,n+d,d) \to \CM(r,n)$.
 Then the following holds. 
\begin{theo}\label{relquotquiver} 
 There is an isomorphism $\CM_D(r,n,d) \simeq \CN(r,n+d,d) $ 
 of schemes over $\CM(r,n)$. 
 \end{theo}
 
 {\it Proof.} The proof relies on the Beilinson spectral sequence, by analogy with 
 the proof of the ADHM correspondence \cite[Thm 2.1]{hilblect}. 
Detailed computations has been carried out in a similar context in 
\cite[Sect. 7.1]{modADHM},\cite{Henni}, therefore it suffices here to outline the 
 main steps, omitting many details.

Now recall that the Beilinson spectral sequence yields an isomorphism 
\cite[Thm 2.1]{hilblect} 
between  the moduli stack of framed torsion free sheaves on $S$ 
with fixed numerical invariants $(r,n)$ and a 
moduli stack of three-term locally free monad complexes on $S$. 
The same correspondence exists for families of framed sheaves; this has been worked out in \cite{Henni} when $S$ is a blowup of the complex plane, but it can be easily adapted to the case of $\IP^2$.
More specifically, let $(E_T,\xi_T)$ be a flat family of framed torsion 
free sheaves on $S$ parameterized by a scheme $T$ of finite type over $\IC$. 
Let $p_T:T\times S \to T$, $p_S:T\times S\to S$ denote the canonical projections and for any coherent sheaf $F_T$ on $T\times S$, $F_T(k) = F_T \otimes p_S^*\CO_S(k)$ for any $k\in \IZ$. 
One can check that the direct images $R^ip_{T*}(E_T(-1))$ vanish for $i=0,2$, 
and $R^1p_{T*}(E_T(-1))$ is a locally free sheaf $\CV_{T}$ of rank $n$ on $T$.
Then the relative Beilinson spectral  spectral sequence for the projective bundle 
$T\times S \to T$ collapses to 
a monad complex $F(E_T,\xi_T)$ of the form 
\be\label{eq:monadA} 
 p_T^*\CV_{T} (-1) {\buildrel a_T\over \longto} p_T^*\CV_T^{\oplus 2}\oplus p_T^*\CW_T{\buildrel b_T\over\longto} p_T^*\CV_T(1).
\ee
where 
\[
\CW_T=R^0p_{T*}E\otimes\CO_{T\times D_\infty} \simeq \CO_T^{\oplus r}.
\]
The differentials $a_T,b_T$ are of the form 
\[ 
\bal 
a_T = \left[\begin{array}{c} z_1-z_0A_{T,1}\\ z_2-z_0A_{T,2} \\ 
z_0J_T\\ \end{array}\right]\qquad 
b_T = [\ -z_2 + z_2A_{T,2} & 
z_1-z_0A_{T,1} & z_0I_T \ ] 
\eal 
\]
where 
\[
(A_{T,1}, A_{T,2}, I_{T}, J_T)\in \mathrm{End}(\CV_T)^{\oplus 2}\oplus 
\mathrm{Hom}(\CW_T, \CV_T) \oplus \mathrm{Hom}(\CV_T, \CW_T)
\]
is a flat family of stable representations of the ADHM quiver. 
Recall that the monad complex $F(E_T,\xi_T)$ is exact at both ends, 
and its middle cohomology sheaf is isomorphic to $E_T$. The three terms have degrees $0,1,2$ respectively. Recall also that 
\[
I_T : \CW_T = R^0p_{T*}E\otimes\CO_{T\times D_\infty} \to \CV_T = 
R^1p_{T*}E(-1)
\]
is the natural coboundary morphism. 

There is a similar isomorphism between the moduli stack of degree $d$ skyscraper sheaves $G$ on $D$ with support disjoint from $D_\infty$ and a moduli 
stack of locally-free two-term complexes on $D=\IP^1$.  
Given a flat family $G_T$ of such objects parameterized by a scheme $T$, the 
corresponding two-term monad complex $F(G_T)$ is 
\be\label{eq:monadB} 
p_T^*\CV_{T,2}(-1) {\buildrel b_{T,2}\over \longto} p_T^*\CV_{T,2}
\ee
where $\CV_{T,2} = R^0p_{T*}G_T$ is a locally free $\CO_T$-module, 
and the terms have degrees $-1,0$ respectively. The 
differential is of the form 
\[
b_{T2} = [\ z_2-z_0B_{T,2}\ ] 
\]
where $B_{T,2} \in \mathrm{End}(\CV_{T,2})$ is an endomorphism 
of $\CV_{T,2}$. 

Let $g_T :E_T\to G_T$ be a surjective morphism of $\CO_{T\times S}$-modules, 
and let $\tilde E_T=\mathrm{Ker}(g_T)$; $\tilde E_T$ is a flat family of 
torsion free $\CO_{S}$-modules. Since the support of $G_T$ is disjoint 
from $T\times D$, there is a canonical isomorphism $E_T\otimes
\CO_{T\times D_\infty}\simeq \tilde E_T\otimes \CO_{T\times D_\infty}$. 
Therefore the framing of $E_T$ along $T\times D_\infty$ yields a 
framing $\xi'_T$ of $\tilde E_T$. 
Therefore the Beilinson spectral sequence of $\tilde E_T$
is again a monad complex $\CF(\tilde E_T,\xi_T')$. Let $\CV_{T,1}=R^1p_{T*}
\tilde E_T$.  
Since the Beilinson spectral sequence is functorial, the exact sequence 
\be\label{eq:exseqsheaves}
0\to \tilde E_T \to E_T \to G_T \to 0
\ee
yields an exact triangle of the form 
 \be\label{eq:exmonadseq}
 \CF(G_T)[-1] {\buildrel \varphi\over \longto} \CF(\tilde E_T,\xi_T') \to \CF(E_T,\xi_T). 
 \ee
Proceeding by analogy with  \cite[Sect. 7.1]{modADHM}, it follows that the morphism 
$\varphi:\CF(G_T)[-1] \to \CF(\tilde E_T,\xi_T')$ is a morphism of monad complexes determined by the natural injective morphism of sheaves 
\[
f_T:\CV_{T,2}= R^0p_{T*}G_T\to \CV_{T,1}= R^1p_{T*}\tilde E(-1), 
\]
which satisfies 
\be\label{eq:Trelations}
A_{T,1}f_T=0,\qquad A_{T,2}f_T=f_TB_{2,T}, \qquad J_Tf_T=0. 
\ee
The details  are very similar to those in loc.cit., hence will be omitted. 
In conclusion, there is a morphism of stacks between the stack of data 
$((E,\xi), G,g)$ on $S$ and the moduli stack of  stable framed representations 
of the enhanced ADHM quiver. 

Conversely, suppose $\CR_T=(\CV_{T,1}, \CV_{T,2}, A_{T,1}, A_{T,2}, I_T, J_T, B_{T,2}, f_{T})$ is a flat family of  stable framed quiver representations parameterized by $T$ with $\CW_T = \CO_{T}^{\oplus r}$. Since the relations 
\eqref{eq:Trelations} are satisfied and $\mathrm{Im}(f_T)\cap \mathrm{Im}(I_T)=0$, the data $(A_{T,1}, A_{T,2}, I_T, $ $J_T)$ induce
ADHM data $({\widetilde A}_{T,1}, {\widetilde  A}_{T,2}, {\widetilde  I}_T, 
{\widetilde  J}_T)$ on the quotient sheaf $\CV_{T,1}/\mathrm{Im}(f_T)$
as in lemma (\ref{projlemma}).  Note that this quotient is 
locally free since the restriction of $f_T$ to any point $t\in T$ is injective. 
Moreover, it is straightforward to check that the resulting flat family of ADHM data is a 
flat family of stable ADHM data. Given this data, one can easily construct an exact sequence of monad complexes of the form \eqref{eq:exmonadseq}, which in turns yields an exact sequence of framed shaves of the form \eqref{eq:exseqsheaves}.

\hfill $\Box$

\section{The Quiver Partition Function}\label{sectionfour} 
Summarizing the results obtained so far, a quiver quantum mechanical model 
for BPS states bound to surface operators has been constructed in section 
(\ref{sectiontwo}). The geometry of the moduli space of supersymmetric vacua
 has been studied in detail in section (\ref{sectionthree}). In particular, according to 
Theorems (\ref{flatdirthm}), (\ref{smoothmoduli}), 
in a special chamber in the space of FI parameters, 
the moduli space $\CN(r,n_1,n_2)$ is a virtually smooth quasi-projective 
variety. An important application of these results is a rigorous mathematical 
construction of a counting function for such BPS states, which is the main focus 
of this section. 

From a physics point of view, the BPS counting function is the Witten index of the 
supersymmetric quantum mechanics obtained in section (\ref{sectiontwo}). 
This index can be computed exactly in the Born-Oppenheimer low energy approximation. 
In this limit the gauged linear quantum mechanical model reduces to a one dimensional 
sigma model on the moduli space of supersymmetric vacua, by analogy with 
the two dimensional situation \cite{twodphases}. 
A complete description of this one dimensional sigma model requires an explicit 
computation of the space of fermion zero modes, at any point in the moduli space. 
The zero modes of the fermionic components of chiral 
multiplets are in one-to-one correspondence with the zero modes of the 
bosonic components, by supersymmetry. The zero modes of the fermionic 
components of Fermi multiplets are determined by a system of linear equations 
following from the Yukawa couplings \eqref{eq:yukawaAi}, \eqref{eq:yukawaAii}. A slightly tedious linear algebra 
computation shows that in the special stability chamber 
all these fermionic fields are in fact massive at any point 
in the moduli space. Therefore the only fermion zero  modes in the 
low energy effective action belong to the chiral multiplets. 
By supersymmetry, they must take values in the 
holomorphic tangent space to the moduli space. 
In particular, there are no fermion zero modes with values in the anti-holomorphic tangent 
bundle. This implies that the supersymmetric ground states are in one-to-one 
correspondence with cohomology classes in $\oplus_{i} H^{0,i}(\CN(r,n_1,n_2))$.  
In conclusion, the Witten index 
is given in this case by the virtual holomorphic Euler character $\chi(\CO_{\CN(r,n_1,n_2)})$ 
of the trivial line bundle on the moduli space $\CN(r,n_1,n_2)$.  

Since the moduli space is non-compact, this Euler characteristic is ill-defined, as the cohomology groups are infinite dimensional. However, in instanton computations one is interested 
in an equivariant virtual Euler characteristic with respect to a natural torus action on the 
moduli space \cite{Nekrasov:2002qd}. 
In this case, ${\bf T}= \IC^{\times}\times \IC^{\times} 
\times (\IC^\times)^r$ and the action on the moduli space $\CN(r,n_1,n_2)$ is given by 
\be\label{eq:toractA} 
\bal 
(t_1,t_2,z) \times & (V_1,V_2,W,A_1,A_2,I,J,B_2,f) \longto \\
& (V_1,V_2,W,t_1A_1,t_2A_2,Iz^{-1},zt_1t_2J,t_2B_2,f) \\
\eal
\ee
 where $z=(z_1,\ldots, z_r)\in (\IC^\times)^r$. 
From the point of view of (topologically twisted) supersymmetric quantum mechanics, the equivariant 
Euler characteristic can still be interpreted as an Witten index employing a deformation
of the nilpotent BRST operator \cite{Bruzzo:2002xf}. This solves the non-compactness problem 
because a direct application of a standard fixed point theorem shows that the equivariant Euler character   
is an element of the quotient field of the representation ring of ${\bf T}$.

Finally, note that there is in fact a natural family of equivariant partition functions 
depending on two integers $(p_1,p_2)\in \IZ^2$. These are obtained by coupling the 
quantum mechanical system with a line bundle on $\CN(r,n_1,n_2)$
as  in \cite{Tachikawa:2004ur}. Since $\CN(r,n_1,n_2)$ is a fine moduli space of quiver 
representations, 
it is equipped with a 
universal locally free quiver sheaf. In particular there are three 
tautological bundles $\CV_1,\CV_2,\CW$ on the moduli space 
corresponding to the nodes $e_1,e_2,e_\infty$ of the enhanced ADHM quiver. 
By construction, $\CW\simeq \CO_{\CN(r,n_1,n_2)}^{\oplus r}$. 
Let $\CL_1=\mathrm{det}(\CV_1)$, 
$\CL_2=\mathrm{det}(\CV_2)$  be the determinant 
line bundles of $\CV_1, \CV_2$. For any pair of integers 
$(p_1,p_2)\in \IZ^2$ let $\CL_{(p_1,p_2)}= 
\CL_1^{\otimes p_1}\otimes \CL_2^{\otimes p_2}$.
Then the  partition function of 
 the quantum mechanical system coupled to the line bundle $\CL_{(p_1,p_2)}$
is the equivariant virtual Euler characteristic $\chi^{\sf vir}_T(\CN(r,n_1,n_2),\CL_{(p_1,p_2)})$. 
Note that $\CL_{(p_1,p_2)}$ has by construction a canonical {\bf T}-linearization. In 
principle one can consider more general partition functions twisting the linearization 
of $\CL_{(p_1,p_2)}$ by an arbitrary irreducible representation $S$ of $T$. 
 Therefore the most general quiver partition function is an  equivariant Euler characteristic
of the form $\chi^{\sf vir}_{\bf T}(\CN(r,n_1,n_2),S\otimes \CL_{(p_1,p_2)})$.

Next let the discrete data $r,d \in \IZ_{>0}$, $(p_1,p_2)\in \IZ_2$ and $S$ be fixed. 
Let $(Q_1,Q_2,R_a)$, $a=1,\ldots,r$, 
 denote the canonical generators of the representation ring 
of ${\bf T}$ and $(q_1,q_2,\rho_a)$, $a=1,\ldots,r$ 
denote their characters. Let $T$ 
be a formal variable. 
Then define a generating function  
\be\label{eq:quivpartfctX} 
\CZ^{(r,d,p_1,p_2,S)}_{quiv}(q_1,q_2,\rho_a,T) = \sum_{n\geq 0} \ch_{\bf T}\, 
\chi^{\sf vir}_{\bf T}(\CN(r,n+d,d),S\otimes \CL_{(p_1,p_2)}) T^n 
\ee
where $\ch_T(R)$ denotes the character of the representation $R$ of {\bf T}.  
A combinatorial formula for this counting function will be derived in the following 
by equivariant localization. This requires an explicit classification 
of the {\bf T}-fixed loci in the moduli space $\CN(r,n+d,d)$, and a computation
of the equivariant normal bundles to the fixed loci. 

\subsection{{\bf T}-fixed loci and nested Young diagrams}\label{sectionfourone}
The {\bf T}-fixed loci in $\CN(r,n+d,d)$
 will be classified in terms of pairs of nested Young diagrams, which are 
 defined as follows. 
 
 Recall that a Young diagram is a finite set $\mu$ of integral points 
 $(i,j)\in (\IR_{\geq 1})^2$ 
 with the property that if $\mu$ contains a point 
 $(i,j) \in (\IR_{\geq 1})^2$, 
 then it contains all integral points $(i',j') \in (\IR_{\geq 1})^2$ so that 
 $1\leq i'\leq i$ and $1\leq j'\leq j$. 
 To fix conventions, the number of columns of  a (nonempty) 
 Young diagram $\mu$ will 
 be denoted by $c_\mu\in \IZ_{\geq 1}$, the columns being labelled by $i=1,\ldots, c_\mu$. The number of rows will be denoted by $l_\mu\in \IZ_{\geq 1}$, the rows 
 being labelled by $j=1,\ldots,l_\mu$. 
 The number of points in the $i$-th column of $\mu$ will be denoted by $\mu_i$. 
 Note that the number of points in the $j$-th row equals the number of points 
 $\mu^t_j$ in the $j$-th column of the transpose diagram $\mu^t$. Obviously, 
 $\mu_i=0$ unless $1\leq i\leq c_\mu$, $h_0\geq h_1\cdots  
  \geq \mu_{c_\mu}$, and $\mu_1+\cdots+\mu_{c_\mu}=|\mu|$.  If $\mu$ is empty, by convention $c_\mu=0$ 
  and $\mu_i=0$ for all $i\in \IZ$. 
   
 A pair $(\mu,\nu)$ of Young diagrams will be called a pair of nested Young diagrams if  
 there is an inclusion  $\nu\subseteq \mu$ so that the complement $\mu\setminus 
 \nu$ satisfies the following condition
 \begin{itemize}
 \item[$(N)$] If $(i,j)\in \mu\setminus \nu$, 
 then $(i+1,j) \notin\mu$.
  \end{itemize}
 
 Ordered sequence of $r\geq 1$ Young diagrams will be denoted by 
  ${\underline \mu}=(\mu^a)_{1\leq a\leq r}$ and $r$ will be called the length of the sequence. 
 The size of the sequence if defined as 
 \[
 |{\underline \mu}|= \sum_{a=1}^r |\mu^a|.
 \]
  
 A pair  
 $(\mu,\nu)$ of ordered sequences of equal length 
 will be called nested if $(\mu^a,\nu^a)$ is a pair of nested Young diagrams 
 for all $1\leq a\leq r$. 
 Given  such a pair $(\mu,\nu)$ of nested sequences, 
  the number of columns of  $\mu^a$, 
 $\nu^a$ will be denoted by $c^a\in \IZ_{\geq 0}$, $e^a\in \IZ_{\geq 0}$ respectively, for $a=1,\ldots,r$.  
 The height of the $i$-th column of $\mu^a$ will 
 be denoted by $\mu^a_{i}$, and the height of the $i$-th column of $\nu^a$ will 
 be denoted by $\nu^a_{i}$, for $a=1,\ldots,r$.  The pair 
 $(|{\underline \mu}|,    |{\underline \nu}|)\in (\IZ_{\geq 0})^2$ 
 will be called the numerical type of the pair of nested sequences.  
 
  Note that condition $(N)$ implies that no two points in the complement $\mu\setminus \nu$ are allowed to be in the same row. Then it is easy to check 
 that the following inequalities must hold   
 \be\label{eq:columncond} 
 0\leq c^a-e^a \leq 1, \qquad 
 0\leq \mu^a_{i}-\nu^a_{i}\leq \nu^a_{i-1}-\nu^a_{i}
 \ee
 for any $a=1,\ldots, r$, and any $i\geq 0$. If any partition $\mu^a$ or 
 $\nu^a$ is empty, by convention, $c^a=0$, respectively $e^a=0$. 
 Recall also that by convention $\mu_{i}^a=0$, $\nu_{i}^a=0$ if $i>c_{}^a$,
 respectively $i>e^a$.  Moreover,  $Q_1,Q_2, R_a$ denote the 
 one dimensional representations of ${\bf T}$ with characters $t_1,t_2,z_a$, 
   $a=1,\ldots,r$, respectively. 
   
   The classification of {\bf T}-fixed loci in $\CN(r,n_+d,d)$ will be facilitated by
   the existence of the projection morphism ${\mathfrak q}: 
   \CN(r,n+d,d)\to \CM(r,n)$ constructed in lemma (\ref{projlemma}). 
   There is an analogous {\bf T}-action on the moduli space $\CM(n,r)$, 
   the fixed loci being classified in \cite{hilblect} for $r=1$, and 
      \cite{instcountA} for all $r\geq 1$.  
   According to \cite[Prop. 2.9]{instcountA}, the fixed locus $\CM(r,n)^{\bf T}$ is a
   finite set of points in one-to-one correspondence with length $r$ sequences 
   ${\underline \nu}=(\nu^a)_{1\leq a\leq r}$ of Young diagrams so that 
   $|{\underline \nu}|=n$. 
   Moreover, according to \cite{Flume:2002az}, 
   \cite[Thm. 4.2]{instcountA}, the tangent space 
   $T_{\underline{\nu}}\CM(r,n)$, regarded as an element of the representation ring 
   of ${\bf T}$, is given 
   by the following formula 
   \be\label{eq:tangentA} 
   \bal 
   T_{\underline{\nu}}\CM(r,n) = \sum_{a,b=1}^r R_a^{-1}R_b 
  \bigg( \sum_{(i,j)\in \nu^a} Q_1^{i-(\nu^b)^t_j} Q_2^{\nu^a_i-j+1} + 
   \sum_{(i,j)\in \nu^b} Q_1^{(\nu^a)^t_j-i+1} Q_2^{j-\nu_i^b} \bigg)
   \eal 
   \ee
   The analogous result for $\CN(r,n+d,d)$ is  given below. 
 
   \begin{prop}\label{fixedpointlemma}
 The {\bf T}-fixed locus $\CN(r,n+d,d)^{\bf T}$ is a finite set of points in one-to-one correspondence with pairs of nested length $r$ sequences 
 $({\underline \mu},{\underline \nu})=(\mu^a,\nu^a)_{1\leq a\leq r}$ of Young diagrams of 
 type $(|{\underline \mu}|,|{\underline \nu}|)=(n+d,n)$. The virtual tangent space to the 
 moduli space at a 
 {\bf T}-fixed point $({\underline \mu},{\underline \nu})$, 
 regarded as an element of the representation 
 ring of ${\bf T}$, is given by the following formula   
 \be\label{eq:tangentB} 
 \bal 
&  T^{\sf vir}_{({\underline \mu},{\underline \nu})} \CN(r,n+d,d)= \\
 & T_{{\underline \nu}} \CM(n,r)+ 
  \sum_{a,b=1}^r \sum_{i=2}^{e^a+1} \sum_{j= 1}^{c^b} 
 \sum_{s=1}^{\mu_{j}^b-\nu_j^b} 
 R_a^{-1}R_b^{}Q_1^{i-j}\big(Q_2^{\mu_{i}^a-\nu_{j}^b-s+1} - 
 Q_2^{\nu_{i-1}^a-\nu_{j}^b-s+1}\big) \\
 & \qquad \qquad \ + \sum_{a,b=1}^r  \sum_{j=1}^{c^b}\sum_{s=1}^{\mu_{j}^b-\nu_{j}^b}  R_a^{-1}R_b^{} Q_1^{-j+1} Q_2^{\mu_{1}^a-\nu_{j}^b-s+1}.\\
  \eal 
 \ee 
 \end{prop}  
   
 {\it Proof.} Using lemma (\ref{projlemma}) the moduli space of stable framed representations $\CN(r,n+d,d)$ can be alternatively characterized as the moduli 
 space of pairs $\CA=(V_1,W,A_1,A_2,I,J)$, $\wCA=(V,W,\wA_1,\wA_2,\wI,\wJ)$ 
 of stable ADHM data of type $(n+d,r)$, $(n,r)$ respectively, and a surjective 
 morphism ${\widetilde f}: V_1\to V$ of ADHM data such that $A_1|_{\mathrm{Ker}({\widetilde f})}$ is 
 identically zero. Then the correspondence between the {\bf T}-fixed loci in 
 $\CN(r,n+d,d)$ 
 and $r$-collections of pairs of nested Young diagrams is a direct consequence of the  classification of {\bf T}-fixed  loci in the moduli spaces of stable ADHM data $\CM(r,n+d)$, $\CM(r,n)$ \cite[Prop. 2.9]{instcountA}.  

In order to prove equation \eqref{eq:tangentB}, 
 recall that  the virtual tangent space at a closed point $[\CR]\in \CN(r,n+d,d)$ 
 is isomorphic to the difference $H^1(\CC(\CR))-H^2(\CC(\CR))$ between the first and second cohomology groups of the  complex $\CC(\CR)$ 
 constructed in theorem (\ref{smoothmoduli}), equation \eqref{eq:defcomplexA}.  
 Moreover, in the proof of theorem (\ref{smoothmoduli}) it has been proven that 
 there is an exact triangle 
 \be\label{eq:extriangleB}
 \CC(\CR)\longto\CC(\CA)\oplus \CC(\CB) \longto \CC(\CA,\CB), 
 \ee
 where $\CA=(V_1,A_1,A_2,I,J)$, $\CB=(V_2,B_2)$ and the complexes 
 $\CC(\CA)$, $\CC(\CB)$, $\CC(\CA,\CB)$, 
  are given in equations \eqref{eq:defcomplexB}, \eqref{eq:defcomplexC},\eqref{eq:defcomplexD}
  respectively. Note that there is a natural {\bf T}-equivariant structure 
  on the restrictions $\CC(\CA)|_{({\underline \mu},{\underline \nu})}$, 
  $\CC(\CB)|_{({\underline \mu},{\underline \nu})}$, 
  $\CC(\CA,\CB)|_{({\underline \mu},{\underline \nu})}$  to a {\bf T}-fixed point $({\underline \mu},{\underline \nu})=(\mu^a,\nu^a)_{1\leq a\leq r}$ induced by the action 
  of {\bf T} on the moduli space. The resulting {\bf T}-equivariant structures are given below.   
   \be\label{eq:equivdefcomplexB} 
 \CC(\CA):\ \  \begin{array}{c}
 \mathrm{End}(V_1)\\
  \end{array} 
 {\buildrel d_0\over \longto} 
 \begin{array}{c} 
Q_1\otimes \mathrm{End}(V_1)\\ 
 \oplus \\
 Q_2\otimes \mathrm{End}(V_1)\\ 
 \oplus \\
  \mathrm{Hom}(W,V_1) \\
 \oplus\\
 Q_1\otimes Q_2 \otimes \mathrm{Hom}(V_1,W) \\
  \end{array}
 {\buildrel d_1\over \longto} 
 \begin{array}{c} 
 Q_1\otimes Q_2 \otimes \mathrm{End}(V_1)^{}\\ 
 \end{array}  
   \ee
   \be\label{eq:equivdefcomplexC} 
 \CC(\CB):\ \ \mathrm{End}(V_2)\ {\buildrel d_0\over \longto} \
 Q_2\otimes\mathrm{End}(V_2) 
 \ee
 \be\label{eq:equivdefcomplexD} 
   \CC(\CA,\CB):\ \ \begin{array}{c} 
  \mathrm{Hom}(V_2,V_1) \\
 \end{array}
 {\buildrel d_0\over \longto} 
 \begin{array}{c} 
  Q_1\otimes \mathrm{Hom}(V_2,V_1)\\
\oplus \\
  Q_2\otimes \mathrm{Hom}(V_2,V_1)\\
\oplus \\
Q_1\otimes Q_2\otimes \mathrm{Hom}(V_2,W)\\ 
\end{array}  
 {\buildrel d_1\over \longto} 
 \begin{array}{c}
 Q_1\otimes Q_2\otimes \mathrm{Hom}(V_2,V_1) \\
 \end{array} 
  \ee
  where $V_1,V_2,W$ have the following expressions in the representation 
  ring of {\bf T} 
  \be\label{eq:equivspaces} 
  V_1 = \sum_{a=1}^r \sum_{(i,j)\in \mu^{a}} R_aQ_1^{1-i}Q_2^{1-j}, \qquad 
  V_2 = \sum_{a=1}^r \sum_{(i,j)\in \mu^a\setminus \nu^a} 
  R_aQ_1^{1-i}Q_2^{1-j}, \qquad 
  W = \sum_{a=1}^r R_a.
  \ee  
   Note that $\CC(\CA)$ is the equivariant deformation complex of the 
   {\bf T}-fixed  ADHM data  $\CA$. The underlying vector space $V\simeq V_1/V_2$ 
   of the quotient ADHM data $\wCA$ has a similar expression, 
   \be\label{eq:equivquot} 
   V = \sum_{a=1}^r \sum_{(i,j)\in \nu^{a}} R_aQ_1^{1-i}Q_2^{1-j}.
    \ee
    Obviously, $V_1=V+V_2$. 
   Then the exact triangle \eqref{eq:extriangleB} yields the following identity in the 
   representation ring of {\bf T} 
   \be\label{eq:equivdefcomplexE} 
   \bal 
   T^{\sf vir}_{({\underline \mu},{\underline \nu})} \CN(n+d,d,r) = & -(1-Q_1)(1-Q_2)V_1^\vee V_1+W^\vee
   V_1+Q_1Q_2V_1^\vee W \\
   & -(1-Q_2)V_2^\vee V_2 \\
   & + (1-Q_1)(1-Q_2)V_2^\vee V_1 -Q_1Q_2 V_2^\vee W\\
   =&\ T_{\underline \nu}\CM(n,r) +(1-Q_2)(Q_1V^\vee V_2 - V_1^\vee V_2) + 
   W^\vee V_2.\\
   \eal 
   \ee
   Next note that 
   \[
   \bal (1-Q_2)V_1^\vee = & \sum_{a=1}^r R_a^{-1}\sum_{(i,j)\in \mu^a} (1-Q_2)Q_1^{i-1}Q_2^{j-1} \\
   =& \ \sum_{a=1}^r \sum_{i=1}^{c^a}\sum_{j=1}^{\mu_i^a} R_a^{-1}Q_1^{i-1}(Q_2^{j-1}-Q_2^{j}) \\ 
   =& \ \sum_{a=1}^r\sum_{i=1}^{c^a} R_a^{-1}Q_1^{i-1}(1-Q_2^{\mu_i^a}).\\
   \eal 
   \] 
   Similarly, 
   \[
   (1-Q_2)V^\vee =   \sum_{a=1}^r\sum_{i=0}^{e^a-1}R_a^{-1} Q_1^i(1-Q_2^{\nu_i^a}).
   \]   
   Moreover, 
   \[
   V_2 = \sum_{a=1}^r \sum_{i=1}^{c^a} \sum_{s=1}^{\mu_i^a-\nu_i^a} R_a Q_1^{1-i} 
   Q_2^{-\nu_i^a-s+1}.\\
   \]
   Therefore,
   \be\label{eq:termone}
   \bal 
     (1-Q_2)Q_1V^\vee V_2 = & \sum_{a,b=1}^r\sum_{i=1}^{e^a}\sum_{l=1}^{c^b} 
     \sum_{s=1}^{\mu^b_l-\nu_l^b} R_a^{-1}R_bQ_1^{i-l+1} (1-Q_2^{\nu_i^a}) 
     Q_2^{-\nu_l^b-s+1}\\
   =&\ \sum_{a,b=1}^r \sum_{i=2}^{e^a+1}\sum_{l=1}^{c^b} 
   \sum_{s=1}^{\mu^b_l-\nu_l^b} R_a^{-1}R_bQ_1^{i-l} Q_2^{-\nu_l^b-s+1} (1-Q_2^{\nu_{i-1}^a}),\\
      \eal 
   \ee
   \be\label{eq:termtwo}
   \bal 
   -(1-Q_2)V_1^\vee V_2 = & -\sum_{a,b=1}^r \sum_{i=1}^{c^a} 
   \sum_{l=1}^{c^b} \sum_{s=1}^{\mu^b_l-\nu_l^b} R_a^{-1}R_bQ_1^{i-l}  
   Q_2^{-\nu_l^b-s+1} (1-Q_2^{\mu_{i}^a}),\\
   \eal 
   \ee
   \be\label{eq:termthree}
   \bal 
   W^\vee V_2 = & \sum_{a,b=1}^r  \sum_{l=1}^{c^b} 
   \sum_{s=1}^{\mu^b_l-\nu_l^b} R_a^{-1}R_bQ_1^{1-l} Q_2^{-\nu^b_l-s+1}.\\
   \eal
   \ee
   Given inequalities \eqref{eq:columncond}, it follows that the sum  over 
   $i=1,\ldots, c^a$ 
   can be written as a sum over $i=1,\ldots, e^a+1$ employing the convention that 
   $h_{i}^a=0$ for $i\geq c^a+1$. Then \eqref{eq:tangentB} follows from \eqref{eq:equivdefcomplexE} adding the right hand sides of equations 
   \eqref{eq:termone}-\eqref{eq:termtwo}.
   
 \hfill $\Box$
 
 \subsection{Equivariant virtual Euler character }\label{sectionfourtwo}
 Given Proposition (\ref{fixedpointlemma}), the computation of the 
 equivariant Euler character $\chi^{\sf vir}_{\bf T}(\CN(r,n_1,n_2),
 S\otimes \CL_{(p_1,p_2)})$ is a 
 straightforward exercise. Explicit formulas will be given below only for 
 $(p_1,p_2)=(0,1)$, which is the relevant case for comparison with toric 
 open string invariants. For simplicity, let $\CL$ denote $\CL_{(0,1)}$ below. 
    Note that the restriction of ${\CL}$ to the {\bf T}-fixed point 
 $({\underline \mu},{\underline \nu})$ is given by 
 \be\label{eq:equivdetA} 
 {\CL}_{({\underline \mu},{\underline \nu})} =  \prod_{a=1}^r 
 \prod_{i=1}^{c^a}\prod_{s=1}^{\mu^a_i-\nu^a_i} R_a Q_1^{1-i}
 Q_2^{-\nu^a_i-s+1}.
 \ee
    Then the virtual localization theorem yields the following formula for the equivariant 
  Euler character of ${\CL}$. 
  \be\label{eq:equiveulerA} 
  \bal
 \ch_{\bf T} (\chi^{\sf vir}_{\bf T}({\CL})) = &\ 
  \sum_{\substack{({\underline \mu},{\underline \nu})\\ 
  (|{\underline \mu}|,|{\underline \nu}|)=(n+d,n)\\}}  
  {\ch_{\bf T}
  ({\CL}_{({\underline \mu},{\underline \nu})}) \over 
  \Lambda_{-1}(T^{\sf vir}_{({\underline \mu},{\underline \nu})}\CN(r,n+d,d)^\vee)}\\
  =&\ \sum_{\substack{({\underline \mu},{\underline \nu})\\ 
  (|{\underline \mu}|,|{\underline \nu}|)=(n+d,n)\\}}
  {\CW_{({\underline\mu}, {\underline \nu})}(q_1,q_2,\rho_a)\over 
  \Lambda_{-1}(T_{\underline {\nu}}
  \CM(r,n)^\vee)}, \\
  \eal 
  \ee
  where 
  \be\label{eq:equiveulerD}
  \bal 
  \CW_{({\underline\mu}, {\underline \nu})}(q_1,q_2,\rho_a) = & \ 
  {\prod_{a=1}^r 
 \prod_{i=1}^{c^a}\prod_{s=1}^{\mu^a_i-\nu^a_i} \rho_a q_1^{1-i}
 q_2^{-\nu^a_i-s+1}\over \prod_{a,b=1}^r \prod_{i=2}^{e^a+1} \prod_{j=1}^{c^b} 
 \prod_{s=1}^{\mu^b_j-\nu^b_j} (1-\rho_a^{}\rho_b^{-1}q_1^{j-i}q_2^{\nu_j^b+s-\mu_i^a-1})
 }\\  
 & \ {\prod_{a,b=1}^r \prod_{i=2}^{e^a+1} \prod_{j=1}^{c^b} 
 \prod_{s=1}^{\mu^b_j-\nu^b_j} (1-\rho_a^{}\rho_b^{-1}q_1^{j-i}q_2^{\nu_j^b+s-\nu_{i-1}^a-1})
 \over \prod_{a,b=1}^r \prod_{j=1}^{c^b}\prod_{s=1}^{\mu^b_j-\nu^b_j} 
 (1-\rho_a\rho_b^{-1}q_1^{j-1}q_2^{\nu_j^b+s-\mu_1^a-1})},
 \\
   \eal
  \ee
  \be\label{eq:equiveulerB} 
  \bal 
 & {1\over \Lambda_{-1}(T_{{\underline {\nu}}}
  \CM(r,n)^\vee)} = \\
  & {1\over \prod_{a,b=1}^r \prod_{(i,j)\in \nu^a} 
  (1-\rho_a\rho_b^{-1} q_1^{(\nu^b)^t_j-i}q_2^{j-\nu_i^a-1})
   \prod_{(i,j)\in \nu^b} 
  (1-\rho_a\rho_b^{-1} q_1^{i-(\nu^a)^t_j-1}q_2^{\nu_i^b-j})}
  \eal 
  \ee
and 
\[
q_1=\ch_T(Q_1), \qquad q_2=\ch_2(Q_2), \qquad \rho_a = \ch_{\bf T}(R_a), \
a=1,\ldots,r.
\]
Given any collection of $r$ Young diagrams ${\underline \nu}$ and an positive integer $d\in \IZ_{\geq 1}$, 
set 
\be\label{eq:equiveulerE} 
\CW_{{\underline \nu},d}(q_1,q_2,\rho_a) = 
\sum_{\substack{({\underline \mu},{\underline \nu})\\ 
  |{\underline \mu}|=|{\underline \nu}|+d\\}}  \CW_{({\underline\mu}, {\underline \nu})}(q_1,q_2,\rho_a). 
  \ee
 where the sum is over all nested sequences $({\underline \mu},{\underline \nu})$
 of $r$ Young diagrams with fixed 
 ${\underline \nu}$. 
Then, obviously, 
\[
\ch_{\bf T} \chi^{\sf vir}_{\bf T}(\CL) = \sum_{{\underline \nu}} {1\over \Lambda_{-1} 
(T_{\underline {\nu}}
  \CM(r,n)^\vee)} \CW_{{\underline \nu},d}(q_1,q_2,\rho_a).
  \]
In conclusion, for  fixed $r,d\in \IZ_{\geq 1}$, $(p_1,p_2)=(0,1)$ and 
$S$, the quiver partition function \eqref{eq:quivpartfctX} is given by 
\be\label{eq:eulergenfctA} 
\CZ_{quiv}^{(r,d,S)}(q_1,q_2,\rho_a,T) = \sum_{n} T^n \ch_{\bf T}(S) 
\sum_{|{\underline \nu}|=n} {1\over \Lambda_{-1} 
(T_{\underline {\nu}}
  \CM(r,n)^\vee)} \CW_{{\underline \nu},d}(q_1,q_2,\rho_a).
   \ee

\section{Comparison with refined open string invariants}\label{sectionfive}

The goal of this section is to formulate a precise conjecture relating the 
quiver partition functions \eqref{eq:eulergenfctA} , with $r=1,2$, 
to refined open string 
invariants of special lagrangian branes in toric Calabi-Yau threefolds. 
According to \cite{Alday:2009fs,Dimofte:2010tz}, M5-branes wrapping such cycles 
yield surface operators in the five dimensional gauge theory effective action. 
Therefore a direct comparison between the quiver partition \eqref{eq:eulergenfctA} and
refined open string invariants is an important test for the models constructed in this 
paper. 

Five dimensional pure gauge theories with eight supercharges and gauge group 
$SU(r)$, $r\geq 2$ are engineered by toric Calabi-Yau threefolds constructed as follows. 
Let $Y$ be a resolved conifold geometry, that is the total space of $\CO(-1)\oplus 
\CO(-1)$ over $\IP^1$. Note that the finite group $\Gamma_r$ of $r$-th roots of unity 
 acts fiberwise on $Y$ by 
\[
\omega \times (s_1,s_2) \to (\omega s_1,\omega^{-1}s_2) 
\]
where $\omega=e^{2i\pi/ r}$ and $s_1,s_2$ are linear coordinates 
along the fibers. 
The quotient $Z_0$ is a local Calabi-Yau 
threefold with a curve $X\simeq \IP^1$ of $\IC^2/\Gamma_r$ singularities. 
Let $Z\to Z_0$ 
be the natural crepant resolution; $Z$ is a smooth Calabi-Yau threefold 
containing a reducible exceptional divisor with $r-1$ components $S_1,\ldots, S_{r-1}$. Each component $S_i$ is a geometrically ruled surface over $X$ with 
smooth $\IP^1$-fibers.
One can formally allow $r=1$ in this construction, in which case $\Gamma_r$ is trivial, and $Z\simeq Z_0\simeq Y$.

Note that the threefolds $Z$ are toric, therefore they are equipped with canonical 
symplectic $U(1)^3$ actions. The resulting moment map,
$\rho_Z : Z \to \IR^3$ maps $Z$ surjectively onto its Delzant polytope. 
The boundary of the Delzant polytope consists of a collection of 2-dimensional 
faces linearly embedded in $\IR^3$, which intersect along 1-faces. The 1-faces form 
trivalent a trivalent graph $\Delta_Z$ in $\IR^3$, which is the image of the toric 
skeleton of $Z$ under the moment map $\rho_Z$.   The toric skeleton of $Z$ is the union of all rational holomorphic curves in $Z$, both compact and noncompact, preserved by the 
$U(1)^3$-action. The compact components of the toric skeleton are mapped to finite 
1-faces while the non-compact components are mapped to semi-infinite 1-faces. 

Toric special lagrangian cycles $L\subset Z$ can be constructed applying the methods 
of \cite{mirror-discs}, as in section (\ref{sectiontwoone}). They are essentially classified by their image under the moment  map $\rho_Z$, which has to be a half real line embedded 
in the Delzant polytope of $Z$. There is a special class of cycles $L$ such that $\rho_Z(L)$ 
intersects a 1-face of the graph $\Delta_Z$. These cycles have topology $\IR^2\times S^1$ 
and intersect the toric skeleton of $Z$ along a one dimensional orbit of the canonical 
$U(1)$ action. They are naturally classified in external lagrangian cycles, in which case 
$L$ intersects a non-compact component of the toric skeleton, and 
internal cycles, in which case $L$ $L$ intersects a compact component of the toric skeleton.
Equivalently, $\rho_Z(L)$ intersects a semi-infinite 1-face, respectively a finite 1-face 
of $\Delta_Z$. The lagrangian 
cycles of primary interest in the following will be external cycles as shown below 
for $r=1,2$. 

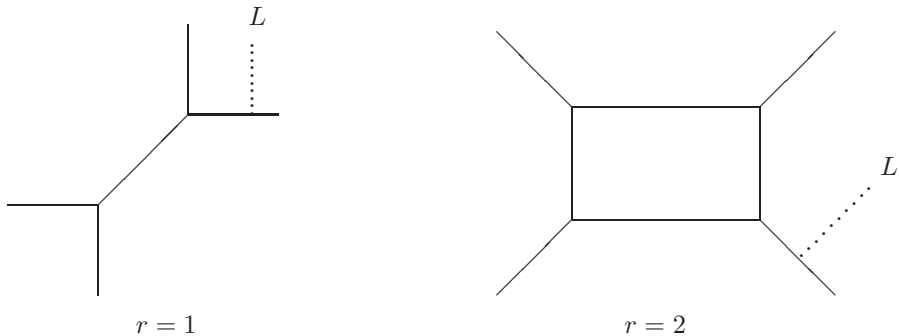
\setlength{\unitlength}{1mm}
\begin{picture}(30,40)
\put(12,12){\line(0,-1){12}}
\put(12,12){\line(-1,0){12}}
\put(12,12){\line(1,1){12}}
\put(24,24){\line(1,0){12}}
\put(24,24){\line(0,1){12}}
\multiput(32,24)(0,1){10}{.}
\put(32,36){$L$}
\put(17,-5){$r=1$}
\put(65,0){\line(1,1){10}}
\put(75,10){\line(1,0){25}}
\put(75,10){\line(0,1){15}}
\put(75,25){\line(-1,1){10}}
\put(100,25){\line(1,1){10}}
\put(100,10){\line(1,-1){10}}
\put(100,10){\line(0,1){15}}
\put(75,25){\line(1,0){25}}
\multiput(105,5)(1,1){10}{.}
\put(116,16){$L$}
\put(82,-5){$r=2$}
%\put(32,11){$\nu_1$}
%\put(49,11){$\nu_1^t$}
%\put(32,26){$\nu_2$}
%\put(49,26){$\nu_2^t$}
%\put(56,22){$\mu^t$}
%\put(56,12){$\mu$}
\end{picture}
\\
\\
\\

The refined open string partition function for an external toric special lagrangian 
cycle $L\subset Z$ 
is constructed using the 
refined topological vertex of \cite{Iqbal:2007ii}, which will be briefly reviewed 
below. 

Given three (possibly empty) Young diagrams $(\lambda,\mu,\nu)$, 
the refined vertex is  a formal series of two variables $(t,q)$ of the form
\be\label{eq:reftopvert} 
\bal
C_{{\lambda}\,{\mu}\,\nu}(t,q)=& \ \Big({t\over q}\Big)^{\frac{||{\mu}||^2}{2}}\,q^{\frac{\kappa({\mu})+||\nu||^2}{2}}\widetilde{Z}_{\nu}(t,q)\\
&\ \sum_{\eta}\Big(\frac{q}{t}\Big)^{\frac{|\eta|+|{\lambda}|-|{\mu}|}{2}}s_{{\lambda}^{t}/\eta}(t^{-\rho}q^{-\nu})s_{{\mu}/\eta}(t^{-\nu^{t}}q^{-\rho})\\
\eal
\ee
where $s_{{\lambda}^{t}/\eta}(t^{-\rho}q^{-\nu})$, $s_{{\mu}/\eta}(t^{-\nu^{t}}q^{-\rho})$ 
are skew Schur functions of the infinite set of variables 
$t^{-\rho}q^{-\nu}=( t^{\frac{1}{2}}q^{-\nu_1}, t^{\frac{3}{2}}q^{-\nu_2},t^{\frac{5}{2}}q^{-\nu_3},\ldots )$
defined in \cite{ORV}, 
\[
{\wZ}_\nu(t,q) = \prod_{(i,j)\in \nu} (1-t^{\nu_j^t-i+1} 
q^{\nu_i-j})^{-1},
\]
and for any partition ${\lambda}$,
%$$n({\lambda})\equiv\sum_i(i-1){\lambda}_i $$
$$|{\lambda}|=\sum_i{\lambda}_i,\qquad ||{\lambda}||=\sum_i{\lambda}_i^2,
\qquad 
\kappa({\lambda})=||\lambda||^2 -||\lambda^t||^2.$$
%$$f_{{\lambda}}(t,q)=(-1)^{|{\lambda}|}\Big(\frac{t}{q}\Big)^{n({\lambda})}q^{-%\frac{\kappa({\lambda})}{2}}.$$
Note that the expression \eqref{eq:reftopvert} differs from \cite[Eqn. 24]{Iqbal:2007ii} 
by the choice of normalization, which is closely related to the normalization chosen in \cite[Sect. 5]{Gukov:2007tf}. Detailed computations will show below that 
\eqref{eq:reftopvert} yields the same results as 
\cite{Iqbal:2007ii} for refined closed string invariants.

The gluing algorithm developed in \cite{Iqbal:2007ii}, assigns to any triple 
$(Z,L,\lambda)$ 
a formal series $\CZ_{\lambda}(q,t,Q)$, which is an expansion in the 
formal variables $Q=(Q_1,\ldots,Q_{M})$
associated to the Mori cone generators of $X$. $\CZ_{\lambda}(q,t,Q)$
is constructed assigning an expression of the form \eqref{eq:reftopvert}
to each trivalent vertex of the dual toric polytope of $Z$, the partitions 
$(\lambda,\mu,\nu)$ being assigned to the edges meeting at the given vertex. 
Then one has to specify gluing rules along edges, eventually including 
certain framing factors, and sum over all partitions associated to finite edges. 
Toric lagrangian branes correspond to infinite edges, and the corresponding 
partitions are not summed over. The details are somewhat intricate and easier 
to explain in concrete examples as shown in sections (\ref{sectionfiveone}), (\ref{sectionfivetwo})
below. 

Suppose there is a stack of $m$ D3-branes wrapped on $L$, the holonomy of 
the flat $U(m)$ gauge field around $S^1$ being in the conjugacy class 
of an element $(\alpha_1,\ldots, \alpha_m)$ of the maximal torus.  
In order to compute the refined 
open topological {\bf A}-model partition function of such a D3-brane system,
let ${y} =
(y_1,y_2,\ldots, )$ be an infinite set of formal variables and let 
\be\label{eq:opentopA} 
\CZ_{open}^{ref}(t,q,Q; {y}) = \sum_{\lambda} \CZ_\lambda(t,q,Q) 
s_\lambda(y) 
\ee
Then the refined open topological partition function of $m$ D3-branes on $L$ 
with holonomy in the conjugacy class of the diagonal matrix $(\alpha_1,\ldots, \alpha_m)$ is obtained by evaluating \eqref{eq:opentopA} at 
${\underline y} = (\alpha_1,\ldots, \alpha_m, 0,0,\ldots)$. 
Note that only Young diagrams $\lambda$ with $|\lambda|\leq m$ contribute to 
this truncation. 

Using this formalism, the quiver partition function \eqref{eq:eulergenfctA} will be
related to the corresponding refined open string partition function 
for $r=1,2$. For $r=1$, the threefold $Z$ is isomorphic to the crepant resolution of a conifold singularity, while for $r=1$, $Z$ is isomorphic to the total space of the 
canonical bundle of $\IP^1\times\IP^1$.

\subsection{Conifold}\label{sectionfiveone}
The resolved conifold is the toric threefold $Y$ isomorphic to the total space of 
$\CO(-1)\oplus \CO(-1)\to \IP^1$. Note that there is only one formal variable $Q$ assigned to the class of the zero section. The toric polytope and projection of special 
lagrangian cycle are represented below. 

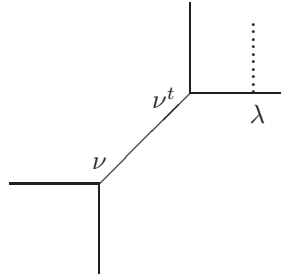
\setlength{\unitlength}{1mm}
\begin{picture}(30,40)
\put(24,12){\line(0,-1){12}}
\put(24,12){\line(-1,0){12}}
\put(24,12){\line(1,1){12}}
\put(36,24){\line(1,0){12}}
\put(36,24){\line(0,1){12}}
\multiput(44,24)(0,1){10}{.}
\put(44,20){$\lambda$}
\put(23,14){$\nu$}
\put(31,22){$\nu^t$}
\end{picture}

Then, applying the refined vertex construction, one obtains 
\be\label{eq:opentopB}
\bal
\CZ_{\lambda}(t,q,Q) = 
\sum_{\nu} & (-Q)^{|\nu|} C_{\emptyset,\emptyset,\nu}(t,q) 
C_{\lambda,\emptyset,\nu^t}(q,t)\\
\sum_{\nu} & (-Q)^{|\nu|}
q^{||\nu||^2/2}t^{||\nu^t||^2/2}
\left({t\over q}\right)^{|{\lambda}|/2}\\
& {\wZ}_\nu(t,q) \wZ_{\nu^t}(q,t) s_{{\lambda}^t}(q^{-\rho}t^{-\nu^t})
\eal
\ee
Note that under the change of variables
\be\label{eq:varchangeA}
t = q_1, \qquad q = q_2^{-1}, \qquad Q = T ({q_1q_2})^{1/2}
\ee
the expression
\[
(-Q)^{|\nu|}
q^{||\nu||^2/2}t^{||\nu^t||^2/2}{\wZ}_\nu(t,q) \wZ_{\nu^t}(q,t)
\]
becomes
\[
{T}^{|\nu|} {1\over  {\Lambda}_{-1} T_\nu(\CM(|\nu|,1)^\vee)}.
\]
Then \eqref{eq:opentopB}
becomes
\be\label{eq:opentopC}
\bal
\CZ_{\lambda}(q_1,q_2^{-1},T(q_1q_2)^{1/2}) = & \sum_{\nu}
{T}^{|\nu|}{1\over {\Lambda}_{-1} T_\nu\CM(|\nu|,1)^\vee} 
\left({q_1q_2}\right)^{|{\lambda}|/2}
s_{{\lambda}^t}(q_2^{-\rho}q_1^{-\nu^t})\\
= & \sum_{\nu} {T}^{|\nu|} {1\over {\Lambda}_{-1} 
T_\nu\CM(|\nu|,1)^\vee}q_1^{|{\lambda}|/2} q_2^{|{\lambda}|}
s_{{\lambda}^t}(q_2^{-1/2}q_2^{-\rho}q_1^{-\nu^t}).\\
\eal
\ee
Redefining the formal variables $y_i$ by 
\[
y_i = q_1^{-1/2}q_2^{-1} x_i 
\] 
for all $i\geq 1$, it follows that 
\be\label{eq:opentopE}
\bal
& \CZ_{open}^{ref}(q_1,q_2^{-1},(q_1q_2)^{1/2}T; q_1^{1/2}q_2x) 
=\\ 
& \sum_{{\lambda}} \sum_{\nu} {T}^{|\nu|} {1\over {\Lambda}_{-1} T_\nu
\CM(|\nu|,1)^\vee}  s_{{\lambda}^t}(q_2^{-1/2}q_2^{-\rho}q_1^{-\nu^t})s_{\lambda}(x)
=\\
& \sum_{\nu} {T}^{|\nu|} {1\over {\Lambda}_{-1} 
T_\nu\CM(|\nu|,1)^\vee}
\prod_{i=1}^\infty \prod_{j=1}^\infty (1 + q_2^{1-i} q_1^{-\nu_i^t} x_j) \\
\eal
\ee
The right hand side of equation \eqref{eq:opentopE} can be expanded in terms 
of the monomial basis $M_\eta({x})$ in the space of symmetric functions, 
which is labelled by partitions $\eta$. Note that for any positive integer $d\in 
\IZ_{>0}$, $M_{(d,0,0,\ldots)}({x}) = x_1^d + x_2^d + \cdots$. 
Let 
$\CZ^{ref}_{open,d}(q_1,q_2,T)$ be the coefficient of $M_{(d,0,0,\ldots)}({x})$ 
in this expansion, which can be computed as follows. 

Let $E_k(x)$, $k\in \IZ_{\geq 0}$ be the degree 
$k$ elementary symmetric function in the variables $x=(x_1,x_2,\ldots)$. 
Then 
\[
\bal 
\mathrm{ln}\, \prod_{i=1}^\infty \prod_{j=1}^\infty (1 + q_2^{1-i} q_1^{-\nu_i^t} x_j) = &\ \mathrm{ln}\, 
\prod_{i=1}^\infty \left(\sum_{k=0}^\infty  q_2^{k(1-i)} q_1^{-k\nu_i^t} 
E_k(x)\right).\\
=&\ \sum_{i=1}^\infty \mathrm{\ln}\, 
\left(\sum_{k=0}^\infty  q_2^{k(1-i)} q_1^{-k\nu_i^t} 
E_k(x)\right)\\
=& \ \sum_{i=1}^\infty \sum_{l=1}^\infty {(-1)^{l-1}\over l} 
\left(\sum_{k=1}^\infty  q_2^{k(1-i)} q_1^{-k\nu_i^t} 
E_k(x)\right)^{l}.\\
\eal
\]
Therefore 
\be\label{eq:expformA}
\prod_{i=1}^\infty \prod_{j=1}^\infty (1 + q_2^{1-i} q_1^{-\nu_i^t} x_j) = 
\mathrm{exp}\, \left[\sum_{i=1}^\infty \sum_{l=1}^\infty {(-1)^{l-1}\over l} 
\left(\sum_{k=1}^\infty  q_2^{k(1-i)} q_1^{-k\nu_i^t} 
E_k(x)\right)^{l}\right]
\ee
In order to compute the coefficients of $M_{(d,0,0,\ldots)}(x) = x_1^d + x_2^d +\cdots $ in the expansion, it suffices to truncate the argument of the exponential function in right hand side of 
\eqref{eq:expformA} to $k=1$ terms, 
\[ 
\mathrm{exp}\, \left[\sum_{i=1}^\infty \sum_{l=1}^\infty {(-1)^{l-1}\over l} 
\left(  q_2^{1-i} q_1^{-\nu_i^t} 
E_1(x)\right)^{l}\right].
\]
Let 
\[
\bal 
F_\nu(q_1,q_2) = &\ \sum_{i=1}^\infty q_2^{1-i} q_1^{-\nu_i^t} = \sum_{i=1}^{l_\nu} q_2^{1-i}q_1^{-\nu_i^t} + {q_2^{-l_\nu}\over 1-q_2^{-1}}.\\
 \eal
\]
Then one has to identify the coefficients of the monomial functions $M_{(d,0,0,\ldots)}(x)$ in the expansion of 
\[
\mathrm{exp}\, \left[\sum_{l=1}^\infty {(-1)^{l-1}\over l} 
E_1(x)^{l}F_\nu(q_1^l,q_2^l)\right], 
\]
which is the same as the coefficient of $x_1^d$ in the expansion of 
\[
\mathrm{exp}\, \left[\sum_{l=1}^\infty {(-1)^{l-1}\over l} 
x_1^l F_\nu(q_1^l,q_2^l)\right].
\]
Expanding the exponential function and collecting all relevant terms, it follows 
that the coefficient of $M_{(d,0,0,\ldots)}(x)$, $d\geq 1$ is 
\be\label{eq:monomcoeffA} 
\bal 
{1\over d!} \sum_{\eta=(1^{d_1}, 2^{d_2}, \ldots)}
{d!\over \prod_{k=1}^d d_k!} \prod_{k=1}^d 
\left({(-1)^{k-1}\over k} 
F_\nu(q_1^{k}, q_2^{k} )\right)^{d_k}
\eal 
\ee
where the sum is over all partitions 
$\eta=(1^{d_1}, 2^{d_2}, \ldots)$ of $d$.

In conclusion the coefficient of $M_{(d,0,0,\ldots)}(x)$ in the right hand side 
of \eqref{eq:opentopE} is 
\be\label{eq:opentopF} 
\bal
\CZ^{ref}_{open,d}(q_1,q_2,T) = \sum_{\nu} {T}^{|\nu|} {1\over {\Lambda}_{-1} 
T_\nu\CM(|\nu|,1)^\vee} \sum_{\eta=(1^{d_1}, 2^{d_2}, \ldots)}
{(-1)^{d-\sum_{k=1}^d d_k}
\over \prod_{k=1}^d (d_k!\, k^{d_k})} \prod_{k=1}^d 
F_\nu(q_1^{k}, q_2^{k} )^{d_k}.
\\
\eal
\ee
For any $\nu$ and $d\geq 1$ let 
\[
\CZ_{\nu,d}(q_1,q_2) = 
\sum_{\eta=(1^{d_1}, 2^{d_2}, \ldots)}
{(-1)^{d-\sum_{k=1}^d d_k}
\over \prod_{k=1}^d (d_k!\, k^{d_k})} \prod_{k=1}^d 
F_\nu(q_1^{k}, q_2^{k} )^{d_k}.
\]
Then the relation between the 
quiver partition function \eqref{eq:eulergenfctA} and the 
refined open topological string partition function \eqref{eq:opentopC} 
is given by:
\begin{conj}\label{quivrefopenconjA}
The following identity holds for any Young diagram $\nu$ and any 
$d\in \IZ_{\geq 1}$. 
\be\label{eq:quivrefopenA} 
\CW_{\nu,d}(q_1,q_2) = \CZ_{\nu,d}(q_1,q_2),
\ee
where $\CW_{\nu,d}(q_1,q_2)$ is defined in equation 
\eqref{eq:equiveulerE}.
In particular 
\be\label{eq:quivrefopenB} 
\CZ_{quiv}^{(1,d,1)}(q_1,q_2,T) = \CZ^{ref}_{open,d}(q_1,q_2,T).
\ee
\end{conj} 
Extensive numerical computations show that conjecture (\ref{quivrefopenconjA}) holds for all 
Young diagrams $\nu$ with $|\nu|\leq 10$ and all $1\leq d\leq 10$. 
A sample computation is presented below. 
\begin{exam}\label{sampleA}
Let $\nu=\tableau{1 3}$ and $d=2$. Then 
\[
F_{\nu}(q_1,q_2) = q_1^{-3} + q_1^{-1} q_2^{-1} + q_2^{-2}(1-q_2^{-1})^{-1}
\]
and 
\[
\bal
\CZ_{\nu,2}(q_1,q_2) = & \ {1\over 2}F_{\nu}(q_1,q_2)^2 - 
{1\over 2}F_{\nu}(q_1^2,q_2^2)\\
= & \ 
{q_1^4+q_1^3q_2^2-q_1^3+q_1q_2^3-q_2q_1-q_2^3+q_2+q_2^4-q_2^2\over 
q_1^4q_2^2(1-q_2)(1-q_2^2)}\\
\eal 
\]
The set of all nested pairs $(\mu,\nu)$ with $|\mu|=|\nu|+2$ consists of the four elements 
$(\mu_1,\nu),\ldots,(\mu_4,\nu)$ represented below. 
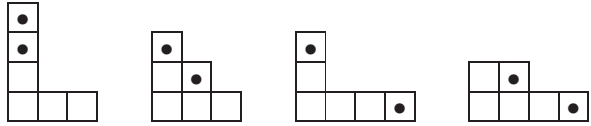
\[
\young(\bullet,\bullet,\hfil,\hfil\hfil\hfil)\qquad 
\young(\bullet,\hfil \bullet,\hfil \hfil\hfil)\qquad 
\young(\bullet,\hfil,\hfil\hfil\hfil\bullet)\qquad 
\young(\hfil\bullet,\hfil\hfil\hfil\bullet)
\]
The boxes in the complement $\mu\setminus \nu$ are marked with $\bullet$. 
Then equation \eqref{eq:equiveulerD} specializes to 
\[ 
\bal 
& \CW_{(\mu_1,\nu), 2}(q_1,q_2) = 
 {q_1^4-q_2q_1-q_1^3+q_2\over q_2^2(1-q_2^2)(1-q_2)(q_1-q_2^2)(q_1^3-q_2^3)}\\
\eal 
\]
\[
\bal 
& \CW_{(\mu_2,\nu), 2}(q_1,q_2) = 
{q_1^5-q_1^3-q_2q_1^2+q_2\over q_1(1-q_2)(q_1^2-q_2)(q_1-q_2^2)(q_1^3-q_2^2)}
\eal 
\]
\[ 
\bal 
& \CW_{(\mu_3,\nu), 2}(q_1,q_2) = 
{(q_1^2+q_2q_1-q_1-q_2)q-2\over q_1^3(q_1^2-q_2)(q_1-q_2)(q_1^2+q_2q_1+q_2^2)(1-q_2)}\eal 
\]
\[ 
\bal 
& \CW_{(\mu_4,\nu), 2}(q_1,q_2) = 
{q_2^2\over q_1^4(q_1-q_2)(q_1^3-q_2^2)}\\
\eal 
\]
Adding the above expressions, it follows that indeed $\CW_{\nu,2}(q_1,q_2)=\CZ_{\nu,2}(q_1,q_2)$.
\end{exam}

\subsection{Local $\IP^1\times\IP^1$}\label{sectionfivetwo}
In this case $Z$ is isomorphic to the total space of the canonical bundle $\CO(-2,-2)$ 
of $\IP^1\times \IP^1$. The Mori cone of $Z$ is generated by the two curve 
classes  associated to the two obvious rulings of $\IP^1\times \IP^1$. 
The corresponding formal variables will be denoted by $Q_f,Q_b$.
The toric polytope and projection of the special lagrangian 
cycle $L$ are represented below. 
\\
\\
\\
\\
\setlength{\unitlength}{1mm}
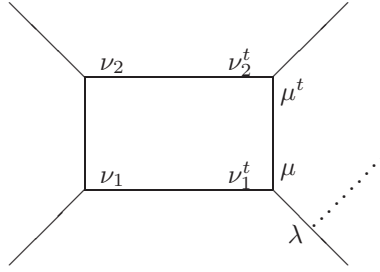
\begin{picture}(30,30)
\put(20,0){\line(1,1){10}}
\put(30,10){\line(1,0){25}}
\put(30,10){\line(0,1){15}}
\put(30,25){\line(-1,1){10}}
\put(55,25){\line(1,1){10}}
\put(55,10){\line(1,-1){10}}
\put(55,10){\line(0,1){15}}
\put(30,25){\line(1,0){25}}
\multiput(60,5)(1,1){10}{.}
\put(57,3){$\lambda$}
\put(32,11){$\nu_1$}
\put(49,11){$\nu_1^t$}
\put(32,26){$\nu_2$}
\put(49,26){$\nu_2^t$}
\put(56,22){$\mu^t$}
\put(56,12){$\mu$}
\end{picture}
\\

By analogy with \cite[Sect. 5.5]{Iqbal:2007ii}, the refined open string partition function 
is 
\be\label{eq:refopA} 
\bal 
\CZ_\lambda(t,q,Q_f,Q_b) = \sum_{\nu_1,\nu_2} (-Q_b)^{|\nu_1|+|\nu_2|} 
{\widetilde f}_{\nu_1^t}(q,t) {\widetilde f}_{\nu_2}(t,q) Z_{\nu_1^t,\nu_2^t,\emptyset}(t,q,Q_f) Z_{\nu_1,\nu_2,\lambda}(q,t,Q_f)
\eal 
\ee
where 
\be\label{eq:refopB} 
\bal 
\CZ_{\nu_1,\nu_2,\lambda}(q,t,Q_f) = 
\sum_{\nu_1,\nu_2,\mu} (-Q_f)^{|\mu|}
C_{\lambda,\mu,\nu_1^t}(q,t) C_{\mu^t,\emptyset,\nu_2^t}(q,t) f_\mu(t,q) 
\eal 
\ee
and 
$f_\eta(t,q)$ ${\widetilde f}_\eta(t,q)$ are framing factors of the form 
\[
f_\eta(t,q) = (-1)^{|\eta|}t^{||\eta^t||^2/2-|\eta|/2} q^{-||\eta||^2/2+|\eta|/2},
\qquad  {\widetilde f}_\eta(t,q) = (-1)^{|\eta|}
\Big({t\over q}\Big)^{|\eta|/2} f_\eta(t,q)
\]

Substituting \eqref{eq:reftopvert} in \eqref{eq:refopB} yields 
\be\label{eq:refopC} 
\bal 
& \CZ_{\nu_1,\nu_2,\lambda}(q,t,Q_f) =\\
&  t^{{||\nu_1^t||^2+||\nu_2^t||^2\over 2}}
\sum_{\nu_1,\nu_2,\mu} Q_f^{|\mu|}\sum_\eta 
\Big({t\over q}\Big)^{{|\eta|+|\lambda|-|\mu|\over 2}}
s_{\lambda^t/\eta}(q^{-\rho}t^{-\nu_1^t})s_{\mu/\eta}(q^{-\nu_1}t^{-\rho})
s_\mu(q^{-\rho}t^{-\nu_2^t}).\\
\eal
\ee
Using the skew Schur function identities 
\[
\sum_\alpha s_{\alpha/\eta_1}(x)s_{\alpha/\eta_2}(y)= \prod_{i,j}
(1-x_i y_j)^{-1} \sum_\kappa
s_{\eta_2/\kappa}(x)s_{\eta_1/\kappa}(y) $$ $$ \sum_\alpha
s_{\alpha^t/\eta_1}(x)s_{\alpha/\eta_2}(y) = \prod_{i,j} (1+x_i
y_j) \sum_\kappa s_{\eta_2^t/\kappa^t}(x)s_{\eta_1^t/\kappa}(y),
\]
it follows that 
\[
\bal 
& \sum_\mu Q_f^{|\mu|}\Big({q\over t}\Big)^{|\mu|/2} 
s_{\mu/\eta}(q^{-\nu_1}t^{-\rho})
s_\mu(q^{-\rho}t^{-\nu_2^t}) = \\
& \prod_{i,j\geq 1}(1-Q_f q^{j-\nu_{1,i}} t^{i-1-\nu_{2,j}^t})^{-1} 
s_\eta(Q_f q^{-\rho+1/2} t^{-\nu_2^t-1/2}) \\
\eal
\]
\[
\bal 
& \sum_{\lambda} \Big({t\over q}\Big)^{|\lambda|/2} 
s_{\lambda^t/\eta}(q^{-\rho}t^{-\nu_1^t})s_\lambda(y) =
& \prod_{i,j\geq 1} (1+q^{i-1} t^{-\nu_{1,j}^t+1/2} y_j) 
s_{\eta^t}(t^{1/2}q^{-1/2}y)\\
\eal
\]
Then 
\be\label{eq:refopD} 
\bal 
& \sum_{\lambda} \CZ_{\nu_1,\nu_2,\lambda}(q,t,Q_f) s_\lambda(y) = \\
& t^{{||\nu_1^t||^2+||\nu_2^t||^2\over 2}} \prod_{i,j\geq 1}(1-Q_f q^{j-\nu_{1,i}} t^{i-1-\nu_{2,j}^t})^{-1} \prod_{i,j\geq 1} (1+q^{i-1} t^{-\nu_{1,j}^t+1/2} y_j) \\
& \sum_\eta \Big({t\over q}\Big)^{|\eta|/2} 
 s_\eta(Q_f q^{-\rho+1/2} t^{-\nu_2^t-1/2})s_{\eta^t}(t^{1/2}q^{-1/2}y) =\\
  & t^{{||\nu_1^t||^2+||\nu_2^t||^2\over 2}} \prod_{i,j\geq 1}(1-Q_f q^{j-\nu_{1,i}} t^{i-1-\nu_{2,j}^t})^{-1} \prod_{i,j\geq 1} (1+q^{i-1} t^{-\nu_{1,j}^t+1/2} y_j) \\
 & \prod_{i,j\geq 1} (1+ Q_f q^{i-1} t^{-\nu_{2,j}^t+1/2}y_j)\\
 \eal 
\ee
Taking into account the framing factors in \eqref{eq:refopA} and redefining 
$y_j=t^{1/2}x_j$, it follows that 
\be\label{eq:refopE} 
\bal 
& \CZ^{ref}_{open}(t,q,Q_f,Q_b;t^{1/2}x) =\\
& \sum_{\nu_1,\nu_2} (-Q_b)^{|\nu_1|+|\nu_2|}q^{||\nu_1||^2} t^{||\nu_2^t||^2} {{\widetilde Z}_{\nu_1}}(t,q) 
{{\widetilde Z}_{\nu_1^t}}(q,t) 
{{\widetilde Z}_{\nu_2}}(t,q) 
{{\widetilde Z}_{\nu_2^t}}(q,t) \\
& \qquad  P_{(\nu_1,\nu_2)}(t,q,Q_f) 
\prod_{i,j\geq 1} (1+q^{i-1} t^{-\nu_{1,j}^t} x_j) 
(1+ Q_f q^{i-1} t^{-\nu_{2,j}^t}x_j)\\
\eal 
\ee
where 
\[
P_{\nu_1,\nu_2}(t,q,Q_f) = 
\prod_{i,j\geq 1}(1-Q_f q^{j-\nu_{1,i}} t^{i-1-\nu_{2,j}^t})^{-1} 
\prod_{i,j\geq 1}(1-Q_f t^{i-\nu_{1,j}^t} q^{j-1-\nu_{2,i}^t})^{-1}
\]
For the purpose of comparison with the quiver partition function, one has to consider the 
normalized partition function ${\widetilde \CZ}^{ref}_{open}(t,q,Q_f,Q_b;t^{1/2}x)$ obtained by replacing $P_{\nu_1,\nu_2}(t,q,Q_f)$ in equation \eqref{eq:refopD} 
by 
\[
{P_{\nu_1,\nu_2}(t,q,Q_f)\over P_{\emptyset,\emptyset}(t,q,Q_f)} = 
\prod_{i,j\geq 1} {(1-Q_ft^{i-1}q^j)(1-Q_fq^{i-1}t^j)\over 
(1-Q_f q^{j-\nu_{1,i}} t^{i-1-\nu_{2,j}^t})
(1-Q_f t^{i-\nu_{1,j}^t} q^{j-1-\nu_{2,i}^t})}.
\]
Proceeding by analogy with \cite[Sect. 5.5.1]{Iqbal:2007ii} it follows that 
\[
\bal 
& {1\over \Lambda_{-1}(T_{\nu_1,\nu_2}\CM(|\nu_1|+|\nu_2|,2)^\vee)}= \\
& q^{||\nu_1||^2} t^{||\nu_2^t||^2} \Big(-Q_f{t\over q}\Big)^{{|\nu_1|+|\nu_2|}}
{{\widetilde Z}_{\nu_1}}(t,q) 
{{\widetilde Z}_{\nu_1^t}}(q,t) 
{{\widetilde Z}_{\nu_2}}(t,q) 
{{\widetilde Z}_{\nu_2^t}}(q,t)
{P_{\nu_1,\nu_2}(t,q,Q_f)\over P_{\emptyset,\emptyset}(t,q,Q_f)} 
\Bigg|_{\substack{t=q_1,\ q=q_2^{-1}\\ Q_f = \rho_1^{-1}\rho_2\\}}\\
\eal
\]
Therefore 
\be\label{eq:refopE} 
\bal 
& {\widetilde \CZ}^{ref}_{open}(q_1,q_2^{-1},\rho_1^{-1}\rho_2,q_1q_2\rho_1^{-1}\rho_2T;t^{1/2}x) =\\
& \sum_{\nu_1,\nu_2}
{T^{|\nu_1|+|\nu_2|} \over \Lambda_{-1}(T_{\nu_1,\nu_2}\CM(|\nu_1|+|\nu_2|,2)^\vee)}
\prod_{i,j\geq 1} (1+q_2^{1-i} q_1^{-\nu_{1,j}^t} x_j) 
(1+ \rho_1^{-1}\rho_2 q_2^{1-i} q_1^{-\nu_{2,j}^t}x_j)\\
\eal 
\ee
Let $\rho_{12} =\rho_1^{-1}\rho_2$. 
Proceeding by analogy with section (\ref{sectionfiveone}), \eqref{eq:opentopE} -- \eqref{eq:opentopF}, the coefficient of 
$M_{(d,0,\ldots)}(x_1,x_2,\ldots)$ in the expansion of the right hand side of 
\eqref{eq:refopE} is 
\be\label{eq:refopF} 
\bal 
& {\widetilde \CZ}^{ref}_{open,d}(q_1,q_2,\rho_{12},T) =\\
& \sum_{\nu_1,\nu_2}
{T^{|\nu_1|+|\nu_2|} \over \Lambda_{-1}(T_{\nu_1,\nu_2}\CM(|\nu_1|+|\nu_2|,2)^\vee)}
\CZ_{(\nu_1,\nu_2),d}(q_1,q_2,\rho_{12})
\eal 
\ee
where 
\[
\bal 
\CZ_{(\nu_1,\nu_2),d}(q_1,q_2,\rho_1^{-1}\rho_2)= 
\sum_{\eta=(1^{d_1}, 2^{d_2}, \ldots)}
{(-1)^{d-\sum_{k=1}^d d_k}
\over \prod_{k=1}^d (d_k!\, k^{d_k})} \prod_{k=1}^d 
F_{(\nu_1,\nu_2)}
(q_1^{k}, q_2^{k}, \rho_{12}^k)^{d_k}
\eal
\]
and 
\[
\bal 
& F_{(\nu_1,\nu_2)}(q_1,q_2,\rho_{12}) = 
F_{\nu_1}(q_1,q_2) + \rho_{12}F_{\nu_2}(q_1,q_2) = \\
&  \sum_{i=1}^{l_{\nu_1}} q_2^{1-i}q_1^{-\nu_{1,i}^t} + {q_2^{-l_{\nu_1}}
\over 1-q_2^{-1}}
+\rho_{12}\Big( \sum_{i=1}^{l_{\nu_2}} q_2^{1-i}q_1^{-\nu_{2,i}^t} + 
{q_2^{-l_{\nu_2}}
\over 1-q_2^{-1}}\Big)\\
\eal 
\]
Then the relation between the 
quiver partition function \eqref{eq:eulergenfctA} and the 
refined open topological string partition function \eqref{eq:refopE} 
is given by:
\begin{conj}\label{quivrefopenconjB}
The following identity holds for any pair of Young diagrams $(\nu_1,\nu_2)$ and any 
$d\in \IZ_{\geq 1}$. 
\be\label{eq:quivrefopenE} 
\rho_1^{-d}\CW_{(\nu_1,\nu_2),d}(q_1,q_2,\rho_1,\rho_2) = 
\CZ_{(\nu_1,\nu_2),d}(q_1,q_2,\rho_{12}),
\ee
where $\CW_{\nu,d}(q_1,q_2,\rho_1,\rho_2)$ is defined in equation 
\eqref{eq:equiveulerE}.
In particular 
\be\label{eq:quivrefopenF} 
\CZ_{quiv}^{(2,d,R_1^{-d})}(q_1,q_2,\rho_1,\rho_2,T) = {\widetilde \CZ}^{ref}_{open,d}(q_1,q_2,\rho_{12},T).
\ee
\end{conj} 
Again, extensive numerical computations show that conjecture (\ref{quivrefopenconjB}) 
holds for all pairs of 
Young diagrams $(\nu_1,\nu_2)$ with $|\nu_1|+ |\nu_2|\leq 10$ and all $1\leq d\leq 10$. 
A sample computation is presented below. 

\begin{exam}\label{sampleB} 
Let $(\nu_1,\nu_2)=(\tableau{2}, \tableau{1})$ and $d=2$. 
Then there are eight sequences of nested pairs $((\mu_1,\mu_2), (\nu_1,\nu_2))$ 
with $|\mu_1|+|\mu_2|= 5$. The partitions $(\mu_1,\mu_2)$ are listed below 
for all these cases. 
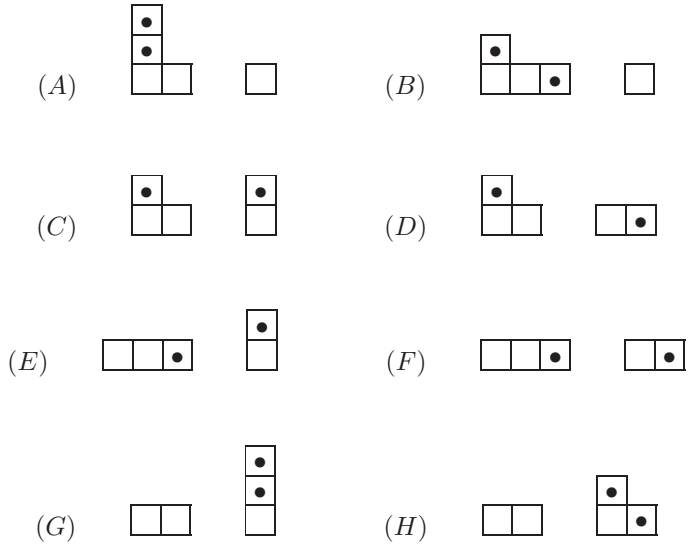
\[
(A) \qquad \young(\bullet,\bullet,\hfil \hfil)\qquad \young(\hfil)\qquad \qquad 
(B) \qquad \young(\bullet,\hfil \hfil \bullet)\qquad \young(\hfil) 
\]
\\
\[
(C)\qquad \young(\bullet,\hfil\hfil)\qquad 
\young(\bullet,\hfil)\qquad \qquad (D)\qquad  \young(\bullet,\hfil\hfil)\qquad \young(\hfil\bullet)
\]
\\
\[
(E)\qquad \young(\hfil\hfil\bullet)\qquad \young(\bullet,\hfil)
\qquad \qquad (F) \qquad 
\young(\hfil\hfil\bullet)\qquad \young(\hfil\bullet)
\]
\\
\[
(G)\qquad \young(\hfil\hfil)\qquad \young(\bullet,\bullet,\hfil)\qquad \qquad 
(H) \qquad \young(\hfil\hfil)\qquad \young(\bullet,\hfil\bullet)
\]
Then 
\[
F_{(\nu_1,\nu_2)}(q_1,q_2,\rho_{12})^2= q_1^{-2}+{q_2^{-1}\over 1-q_2^{-1}} 
+\rho_{12}\left(q_1^{-1}+{q_2^{-1}\over 1-q_2^{-1}}\right )
\]
and 
\[
\bal
& \CZ_{(\nu_1,\nu_2)}(q_1,q_2,\rho_{12}) = {1\over 2}
F_{(\nu_1,\nu_2)}(q_1,q_2,\rho_{12})^2 - {1\over 2} F_{(\nu_1,\nu_2)}(q_1^2,q_2^2,\rho_{12}^2)=\\
& {q_2^2q_1+q_1^3-q_1\over (1-q_2^2)(1-q_2)q_1^3} + \rho_{12} 
{q_2^3+q_1^2q_2^2+q_2^2q_1-q_2^2+q_2q_1^3-q_2+1-q_1^2-q_1+q_1^3\over 
(1-q_2^2)(1-q_2)q_1^3} \\
& + \rho_{12}^2 {q_1^2q_2^2+q_1^3-q_1^2\over 
(1-q_2^2)(1-q_2)q_1^3} \\
\eal 
\] 
Equation \eqref{eq:equiveulerD} specializes respectively to
\[
\CW_{({\underline \mu}, {\underline
\nu}),2}^{(A)}(q_1,q_2,\rho_1,\rho_2)
=\frac{(q_1^2-1)(q_1-\rho_{12})}{(-1+q_2)^2(1+q_2)(q_1^2-q_2^2)(-1+\rho_{12})(-1+q_2\rho_{12})(q_1-q_2^2\rho_{12})}
\]
\[
\CW_{({\underline \mu}, {\underline
\nu}),2}^{(B)}(q_1,q_2,\rho_1,\rho_2)
=\frac{q_2^2(q_1-\rho_{12})(q_2-q_1\rho_{12})}{q_1^2(q_2-1)(q_1^2-q_2^2)(\rho_{12}-1)(q_1\rho_{12}-1)(q_1^2\rho_{12}-q_2)(q_1-q_2\rho_{12})}
\]
\[
\CW_{({\underline \mu}, {\underline
\nu}),2}^{(C)}(q_1,q_2,\rho_1,\rho_2)
=\frac{(q_1-1)^2(1+q_1)q_2^2(q_1-\rho_{12})\rho_{12}^2(q_1^2\rho_{12}-1)}{(q_1-q_2)(q_1^2-q_2)(q_2-1)^2(q_2-\rho_{12})(q_1^2\rho_{12}-q_2)(q_1-q_2\rho_{12})(q_2\rho_{12}-1)}
\]
\[
\CW_{({\underline \mu}, {\underline
\nu}),2}^{(D)}(q_1,q_2,\rho_1,\rho_2)
=-\frac{(q_1^2-1)q_2^2\rho_{12}^2(q_1q_2\rho_{12}-1)}{q_1(q_1-q_2)(q_1^2-q_2)(q_2-1)(\rho_{12}-1)(q_1\rho_{12}-1)(q_1-q_2^2\rho_{12})}
\]
\[
\CW_{({\underline \mu}, {\underline
\nu}),2}^{(E)}(q_1,q_2,\rho_1,\rho_2)
=\frac{(q_1-1)q_2^2\rho_{12}^2(q_1\rho_{12}-q_2)}{q_1^2(q_1-q_2)(q_1^2-q_2)(q_2-1)(\rho_{12}-1)(q_1\rho_{12}-1)(q_1^2\rho_{12}-q_2^2)}
\]
\[
\CW_{({\underline \mu}, {\underline
\nu}),2}^{(F)}(q_1,q_2,\rho_1,\rho_2)
=\frac{q_2^4\rho_{12}^2}{q_1^3(q_1-q_2)(q_1^2-q_2)(q_1^2\rho_{12}-q_2)(q_1-q_2\rho_{12})}
\]
\[
\CW_{({\underline \mu}, {\underline
\nu}),2}^{(G)}(q_1,q_2,\rho_1,\rho_2)
=-\frac{(q_1-1)\rho_{12}^4(q_1^2\rho_{12}-1)}{(q_2-1)^2(q_2+1)(q_2^2-q_1)(q_2-\rho_{12})(\rho_{12}-1)(q_2^2-q_1^2\rho_{12})}
\]
\[
\CW_{({\underline \mu}, {\underline
\nu}),2}^{(H)}(q_1,q_2,\rho_1,\rho_2)
=\frac{q_2^2\rho_{12}^4(q_1^2\rho_{12}-1)(q_1q_2\rho_{12}-1)}{q_1(q_2-1)(q_1-q_2^2)(\rho_{12}-1)(q_1\rho_{12}-1)(q_1^2\rho_{12}-q_2)(q_1-q_2\rho_{12})}
\]
Adding all above expressions confirms identity \eqref{eq:quivrefopenE} in this case. 
\end{exam}

\bibliography{adhmref.bib}
 \bibliographystyle{abbrv}
\end{document}